\documentclass[a4paper,11pt]{article}

\usepackage{jcappub} 
\usepackage{amssymb}
\usepackage{amsthm}

\usepackage{graphicx}

\usepackage[T1]{fontenc} 

\title{Radio measurements of the energy and the depth of the shower maximum of cosmic-ray air showers by Tunka-Rex}

\author[a]{P.A.~Bezyazeekov}
\author[a]{N.M.~Budnev}
\author[a]{O.A.~Gress}
\author[b]{A.~Haungs}
\author[b]{R.~Hiller}
\author[b]{T.~Huege}
\author[a]{Y.~Kazarina}
\author[c]{M.~Kleifges}
\author[a]{E.N.~Konstantinov}
\author[d]{E.E.~Korosteleva}
\author[b]{D.~Kostunin$^{1,}$}
\author[c]{O.~Kr\"omer}
\author[d]{L.A.~Kuzmichev}
\author[d]{N.~Lubsandorzhiev}
\author[a]{R.R.~Mirgazov}
\author[a]{R.~Monkhoev}
\author[a]{A.~Pakhorukov}
\author[a]{L.~Pankov}
\author[d]{V.V.~Prosin}
\author[e]{G.I.~Rubtsov}
\author[b]{F.G.~Schr\"oder$^{1,}$\note{Corresponding authors.}}
\author[f]{R.~Wischnewski}
\author[a]{A.~Zagorodnikov}
\author[]{- Tunka-Rex Collaboration}

\affiliation[a]{Institute of Applied Physics ISU, Irkutsk, Russia}
\affiliation[b]{Institut f\"ur Kernphysik, Karlsruhe Institute of Technology (KIT), Germany}
\affiliation[c]{Institut f\"ur Prozessdatenverarbeitung und Elektronik, Karlsruhe Institute of Technology (KIT), Germany}
\affiliation[d]{Skobeltsyn Institute of Nuclear Physics MSU, Moscow, Russia}
\affiliation[e]{Institute for Nuclear Research of the Russian Academy of Sciences, Moscow, Russia}
\affiliation[f]{DESY, Zeuthen, Germany}

\emailAdd{frank.schroeder@kit.edu}

\abstract{
We reconstructed the energy and the position of the shower maximum of air showers with energies $E \gtrsim 100\,$PeV applying a method using radio measurements performed with Tunka-Rex. 
An event-to-event comparison to air-Cherenkov measurements of the same air showers with the Tunka-133 photomultiplier array confirms that the radio reconstruction works reliably.
The Tunka-Rex reconstruction methods and absolute scales have been tuned on CoREAS simulations and yield energy and $X_{\mathrm{max}}$ values consistent with the Tunka-133 measurements.
The results of two independent measurement seasons agree within statistical uncertainties, which gives additional confidence in the radio reconstruction.
The energy precision of Tunka-Rex is comparable to the Tunka-133 precision of $15\,\%$, and exhibits a $20\,\%$ uncertainty on the absolute scale dominated by the amplitude calibration of the antennas.
For $X_{\mathrm{max}}$, this is the first direct experimental correlation of radio measurements with a different, established method.
At the moment, the $X_{\mathrm{max}}$ resolution of Tunka-Rex is approximately $40\,$g/cm\textsuperscript{2}.
This resolution can probably be improved by deploying additional antennas and by further development of the reconstruction methods, since the present analysis does not yet reveal any principle limitations.
}


\begin{document}
\maketitle
\flushbottom


\section{Introduction}
After more than 100 years of cosmic-ray measurements, the mass composition and the energy spectrum of the primary particles is relatively well-known only in the energy range accessible by direct measurements, $E \lesssim 10^{15}\,$eV. 
At higher energies the flux of cosmic-rays is too low to perform direct measurements with statistical significance. 
Consequently, the knowledge is poorer and relies on the measurement of secondary-particle cascades called air showers. 
These indirect cosmic-ray measurements use the atmosphere as a calorimeter.
The energy of the primary particle can be estimated from the energy contained in the shower. The type of the primary particle can, for example, be deduced from the longitudinal shower development.
Heavy primary particles such as iron nuclei on average interact earlier in the atmosphere than light particles such as protons. 
Thus, the mass composition can be statistically estimated from measurements of the atmospheric depth of the shower maximum, $X_{\mathrm{max}}$.

Established air-shower techniques are the measurement of secondary particles on ground, the measurement of fluorescence or air-Cherenkov light by dedicated telescopes or non-imaging detectors on ground. 
The latter two methods have a relatively high accuracy for the energy and for $X_\mathrm{max}$, but are limited to dark and clear nights. 
Detection of the radio emission by air showers is an additional technique, which does not suffer from this intrinsic limitation of the exposure. 
The radio signal is sensitive to the shower energy~\cite{Allan1971} and $X_{\mathrm{max}}$~\cite{2012ApelLOPES_MTD, 2014ApelLOPES_MassComposition, BuitinkLOFAR_Xmax2014, NellesLOFAR_measuredLDF2015}, too, but it has not yet been demonstrated experimentally that the accuracy for both variables is competitive with those of the other techniques, as indicated by simulation-based studies, e.g., \cite{HuegeUlrichEngel2008, deVries2010, KostuninTheory2015}.

\begin{figure*}
  \centering
  \includegraphics[width=0.7\linewidth]{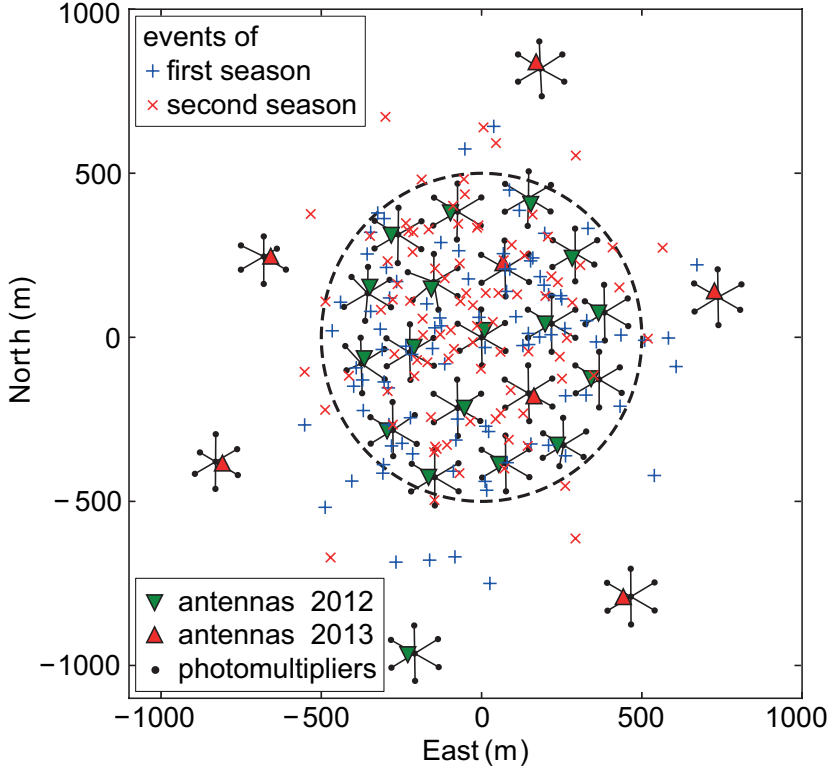}
  \caption{Map of the antenna array Tunka-Rex and its host experiment, the photomultiplier array Tunka-133. 
  18 antennas had been available since the start of the first season in October 2012, and 7 additional antennas have gone in operation only for the second season starting October 2013. 
  The shower cores of the Tunka-Rex events are indicated for both seasons.  
  The dashed circle denotes the cut on the central area with $500\,$m radius used for $X_\mathrm{max}$ reconstruction.}
  \label{fig_map}
\end{figure*}

\begin{figure*}
  \centering
  \includegraphics[width=0.7\linewidth]{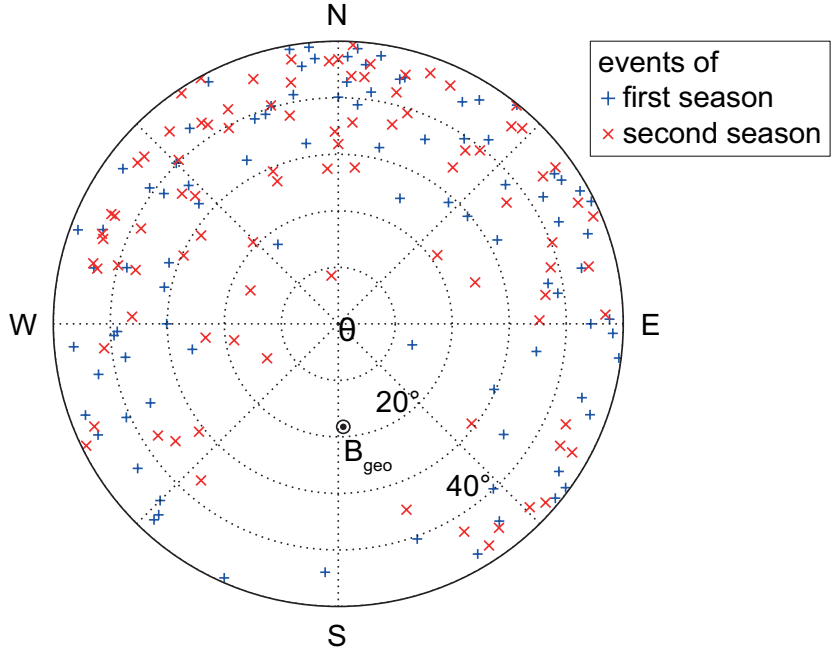}
  \caption{Sky map of the Tunka-Rex events of both seasons. The efficiency rises with zenith angle and is smaller for arrival directions close to the geomagnetic field $\mathbf{B}_\mathrm{geo}$.}
  \label{fig_skyMap}
\end{figure*}

The main origin of the radio emission is the geomagnetic deflection of relativistic electrons and positrons in the shower, which induces a time-variable current~\cite{KahnLerche1966}. 
The amplitude of this emission is proportional to the shower energy and to $\sin \alpha$, with $\alpha$ the angle between the geomagnetic field and the shower direction. 
On a weaker level the Askaryan effect contributes, i.e., radio emission due to the time variation of the net charge excess in the shower front~\cite{Askaryan1962, AugerAERApolarization2014, SchellartLOFARpolarization2014}. 
The interference of both effects leads to an azimuthal asymmetry of the radio footprint~\cite{CODALEMAchargeExcess2015, NellesLOFAR_measuredLDF2015}. 
Finally, the refractive index of air causes Cherenkov-like effects for both emission mechanisms~\cite{AllanICRC1971, Hough1973, deVriesPRL2011}.

Tunka-Rex \cite{TunkaRex_NIM_2015} is the radio extension of the Tunka observatory for cosmic-ray air showers. 
Its main detector, Tunka-133, is an array of non-imaging photomultipliers detecting the Cherenkov light emitted by air-showers in the atmosphere in the energy range of $10^{16}-10^{18}\,$eV. 
Tunka-133 is fully efficient for all zenith angles $\theta \le 50^\circ$ at energies above $10^{16.2}\,$eV~\cite{Tunka133_EPJ2015}, which covers the full energy range of Tunka-Rex.
For Tunka-133 measurements, the energy of the primary particle $E_\mathrm{pr}$ is reconstructed from the flux of the Cherenkov light at $200\,$m distance to the shower axis, and $X_\mathrm{max}$ is reconstructed from the steepness of the amplitude-distance function. 
Using these methods, Tunka-133 features an energy resolution of about $15\,\%$, and an $X_\mathrm{max}$ resolution of about $28\,$g/cm\textsuperscript{2}~\cite{Tunka133_NIM2014}.
For comparison, the average difference between the extreme cases of a pure proton and a pure iron composition is in the order of $100\,$g/cm\textsuperscript{2}. 
Since Tunka-133 triggers Tunka-Rex, the same air showers are measured simultaneously with the air-Cherenkov and the radio detector.

These combined measurements are used for a cross-check of both methods. 
In particular the precision of Tunka-Rex is estimated by a comparison of the energy and $X_\mathrm{max}$ reconstructions to Tunka-133. 
Reconstruction methods for Tunka-Rex have been developed with CoREAS simulations~\cite{HuegeCoREAS_ARENA2012} (the radio extension of CORSIKA \cite{HeckKnappCapdevielle1998}) taking into account previous theoretical predictions~\cite{Kalmykov2012, HuegeUlrichEngel2008, deVries2010, HuegeTheoryOverview_ARENA2012}, and experience by other experiments, in particular LOPES~\cite{2014ApelLOPES_MassComposition}, AERA~\cite{AugerAERApolarization2014} and LOFAR~\cite{SchellartLOFAR2013}. 

The parameters in the reconstruction methods have been determined from the simulations (cf.~reference~\cite{KostuninTheory2015} for details), and not tuned against the Tunka-133 measurements.
First results on energy and $X_\mathrm{max}$ reconstruction have already been presented in reference~\cite{TunkaRexXmaxICRC2015} using the first season of Tunka-Rex measurements from October 2012 to April 2013 (effectively 280 hours of measurements).
The same reconstruction method is now applied also to the second season of data from October 2013 to April 2014 (effectively 260 hours of measurements), and the combined results and the cross-check between both measurement seasons are presented in this article. 
In the experimental data we observe a direct correlation between the Tunka-Rex and Tunka-133 reconstructions, which is used to estimate precision and accuracy of Tunka-Rex for energy and $X_\mathrm{max}$.

\section{Experimental setup and data selection}
Tunka-Rex started in autumn 2012 with 19 antenna stations, and in summer 2013, 6 further antennas went into operation giving a total number of 25 antenna stations used for the present analysis. 
In the central area covering about $1\,$km\textsuperscript{2} the spacing is approximately $200\,$m. The outer antennas contribute only to a few events (see figure~\ref{fig_map}). 
The air-Cherenkov array Tunka-133 covers the inner area with 133 photomultipliers, and features additional photomultipliers in the outer area. 
Upon a coincidence trigger of the photomultipliers, both the radio and the air-Cherenkov detector are read out in parallel. 
The electric field of the radio signal measured by Tunka-Rex is reconstructed in an effective bandwidth of $35-76\,$MHz. 
Details on the detector setup and its calibration can be read in reference~\cite{TunkaRex_NIM_2015}.

The reconstruction of Tunka-133 is fully efficient and has a reliable reconstruction for events with zenith angles $\theta \le 50^\circ$. 
Therefore, only events with $\theta \le 50^\circ$ have been selected for the present analysis (see figure~\ref{fig_skyMap}).
For these events the Tunka-133 reconstruction of the shower provides: energy of the primary particle $E_\mathrm{pr}$, atmospheric depth of the shower maximum $X_\mathrm{max}$, and axis, i.e., the direction and point of incidence on the ground (= shower core). 
For all Tunka-Rex events we used the shower core of Tunka-133 as input, because of the denser spacing of the photomultipliers compared to that of the antennas. 
The direction was only used as cross-check: 
events are considered false-positive, i.e., contaminated by background pulses and excluded from the analysis, if the directions reconstructed from the radio and air-Cherenkov arrival times deviate by more than $5^\circ$~\cite{TunkaRex_NIM_2015}.

For the selection of Tunka-Rex events we applied a quality cut based on the signal-to-noise ratio in individual antenna stations. 
The signal has to exceed a signal-to-noise ratio, $\mathrm{SNR} \equiv (S/N)^2 \ge 10$ in at least three antenna stations, where the signal $S$ is the maximum of an envelope on the electric-field strength of the pulse, and the noise $N$ the root-mean square of the electric field in a time window before the signal. 
This cut is set such that pure background has a chance of about $5\,\%$ to pass in an individual antenna, which balances efficiency and purity. 
Given 25 antenna stations, this leads to a non-negligible probability that in a true event antenna stations with false-positive signals are contained. 
To exclude most of these false-positive signals, we sort the antennas by their distance to the shower axis. 
After two antenna stations failing the SNR cut, any further antenna stations are excluded from the analysis for the particular event. 
Thus, the risk that an event is contaminated by a station with false-positive signal is reduced to approximately $10\,\%$. 
If surviving false-positive signals significantly impact the event reconstruction, they are removed by the cut requiring the $5^\circ$ agreement between the direction reconstructed by Tunka-Rex and Tunka-133. 
Thus, the remaining data set is assumed to be practically free of false-positive signals.
Finally, for the $X_{\mathrm{max}}$ reconstruction we apply additional quality cuts:
At least one antenna above threshold has to be at a distance $r_\mathrm{axis} > 200\,$m to the shower axis, since $X_{\mathrm{max}}$ is reconstructed from the slope of the lateral distribution, which requires a sufficient lever arm. 
Moreover, the shower core has to be inside the central area of the array (dashed circle in figure~\ref{fig_map}), and the estimator for the statistical reconstruction uncertainty of Tunka-Rex must not exceed a certain value: 
$\sigma(X_\mathrm{max}^\mathrm{radio}) \le 50\,$g/cm\textsuperscript{2}.

Reconstruction of the electric-field strengths (= amplitude) in individual antenna stations is performed with a modified version of the `Auger Offline' software developed by the Pierre Auger Collaboration~\cite{RadioOffline2011}. 
This software applies all known experimental characteristics to the recorded signals, in particular the frequency-dependent gain and pulse distortion of the various components in the signal chain as determined by calibration measurements~\cite{TunkaRex_NIM_2015}. 
For the reconstruction of the electric-field vector the simulated direction dependence of the antenna gain is used, which has been validated for a few selected directions by calibration measurements. 

\begin{figure*}[t]
  \centering
  \includegraphics[width=0.6\linewidth]{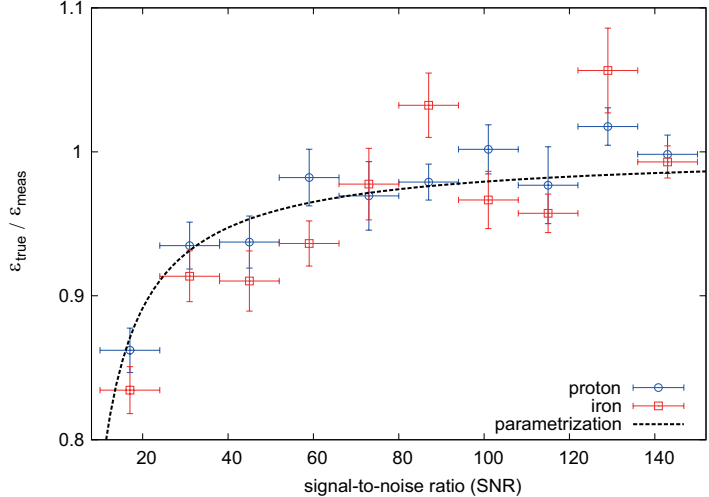}
  \caption{Average ratio of the true pulse amplitude of CoREAS simulations $\mathcal{E}_\mathrm{true}$ and the amplitude after adding measured noise $\mathcal{E}_\mathrm{meas}$ over the signal-to-noise ratio.}
  \label{fig_noiseInfluence}
\end{figure*}

For the suppression of narrow-band radio-frequency interferences (RFI) we apply rectangular band-stop filters of $\pm 0.2\,$MHz width (corresponding to 3 bins) at each integer $5\,$MHz ($35$, $40$, ..., $75\,$MHz), since these frequencies are often contaminated by RFI background 
\footnote{In earlier analyses we used a gliding median filter to suppress narrow-band RFI, but it turned out that this filter significantly increases the impact of background on the measured amplitudes, without significantly increasing the signal-to-noise ratio. 
Nevertheless, the exact method of RFI suppression has only minor influence on the high-quality events used in the present analysis.}.
Since the radio emission from the air shower is broadband, it is significantly less affected by the filter than the narrow-band disturbances. 
The background remaining after filtering still on average increases the measured pulse height $\mathcal{E}_\mathrm{meas}$ compared to the true pulse height $\mathcal{E}_\mathrm{true}$, as already seen at the LOPES experiment~\cite{SchroederNoise2010} and assumed for early experiments~\cite{AllanFreqSpecAndNoise1970}. 
We studied this effect by adding measured background to about 300 CoREAS simulations (figure~\ref{fig_noiseInfluence}), and found that the parametrization given in reference~\cite{SchroederNoise2010} has to be slightly modified introducing a normalization factor $k = 4.1$ adapted for Tunka-Rex. 
This normalization factor reflects the different ways how the noise level is measured in LOPES (mean of local maxima in the instantaneous amplitude) and Tunka-Rex (root mean square of the electric field, which corresponds to the mean power of noise).
Thus, we use the following formula to correct for the average effect of noise:

\begin{equation}
\mathcal{E}_\mathrm{true} = \mathcal{E}_\mathrm{meas} \cdot \sqrt{1 - \frac{k}{\mathrm{SNR}}} \,,
\label{noise_correction}
\end{equation}
with the signal-to-noise ratio $\mathrm{SNR}$ defined above.

\begin{table}[t]
\centering
\caption{Statistics of Tunka-Rex events triggered at zenith angles $\theta \le 50^\circ$ in both seasons after certain quality cuts. 
The cuts are applied consecutively in the given order. 
The event numbers denote the number of surviving events after application of all cuts up to that point.
The first three cuts are used to reject false-positive events contaminated by background pulses. 
The remaining $91 + 87$ events are used for energy reconstruction. 
For $X_{\mathrm{max}}$ reconstruction three additional quality cuts are applied leaving $23 + 19$ events for the two measurement seasons, respectively.} \label{tab_Eventstatistics}
\vspace{0.3cm}
\begin{tabular}{lcc}
\hline
Number of events & first season of & second season of\\
after cut on & Oct. 2012 - Apr. 2013 & Oct. 2013 - Apr. 2014\\
             & (280 effective hours) & (260 effective hours)\\ 
\hline
$\ge 3$ stations with signal & 244 & 445 \\
rejecting outlier stations & 122 & 147 \\
$\measuredangle$ (Tunka-Rex, Tunka-133) $\le 5^\circ$ & \textbf{91} & \textbf{87} \\
at least one antenna at $r_\mathrm{axis} > 200\,$m & 64 & 56 \\
$\sigma(X_\mathrm{max}^\mathrm{radio}) \le 50\,$g/cm\textsuperscript{2} & 25 & 22\\
inner area ($r<500\,$m) & \textbf{23} & \textbf{19}\\
\hline
\end{tabular}
\end{table}

\begin{figure*}
  \centering
  \includegraphics[width=0.45\linewidth]{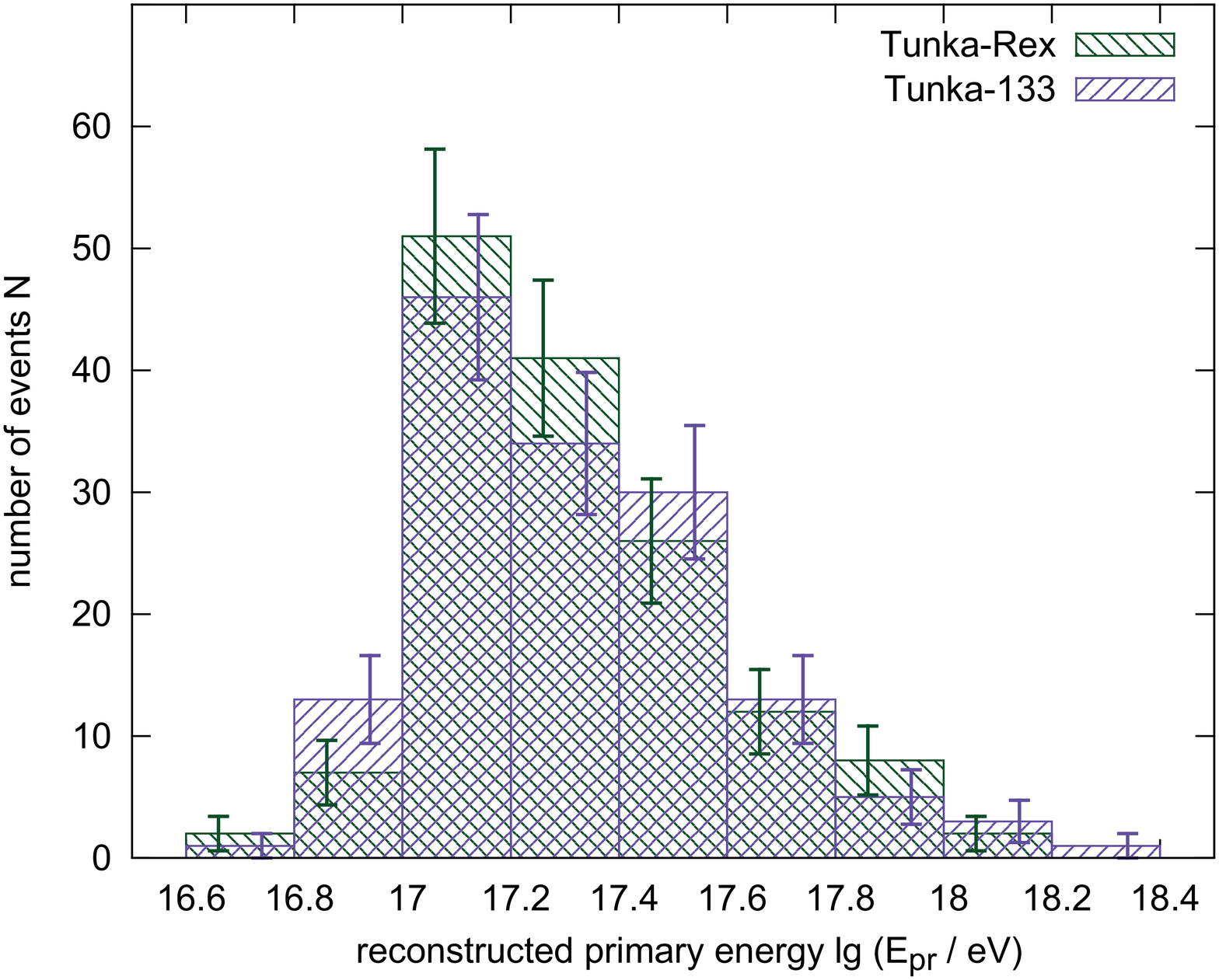}
  \hfill
  \includegraphics[width=0.45\linewidth]{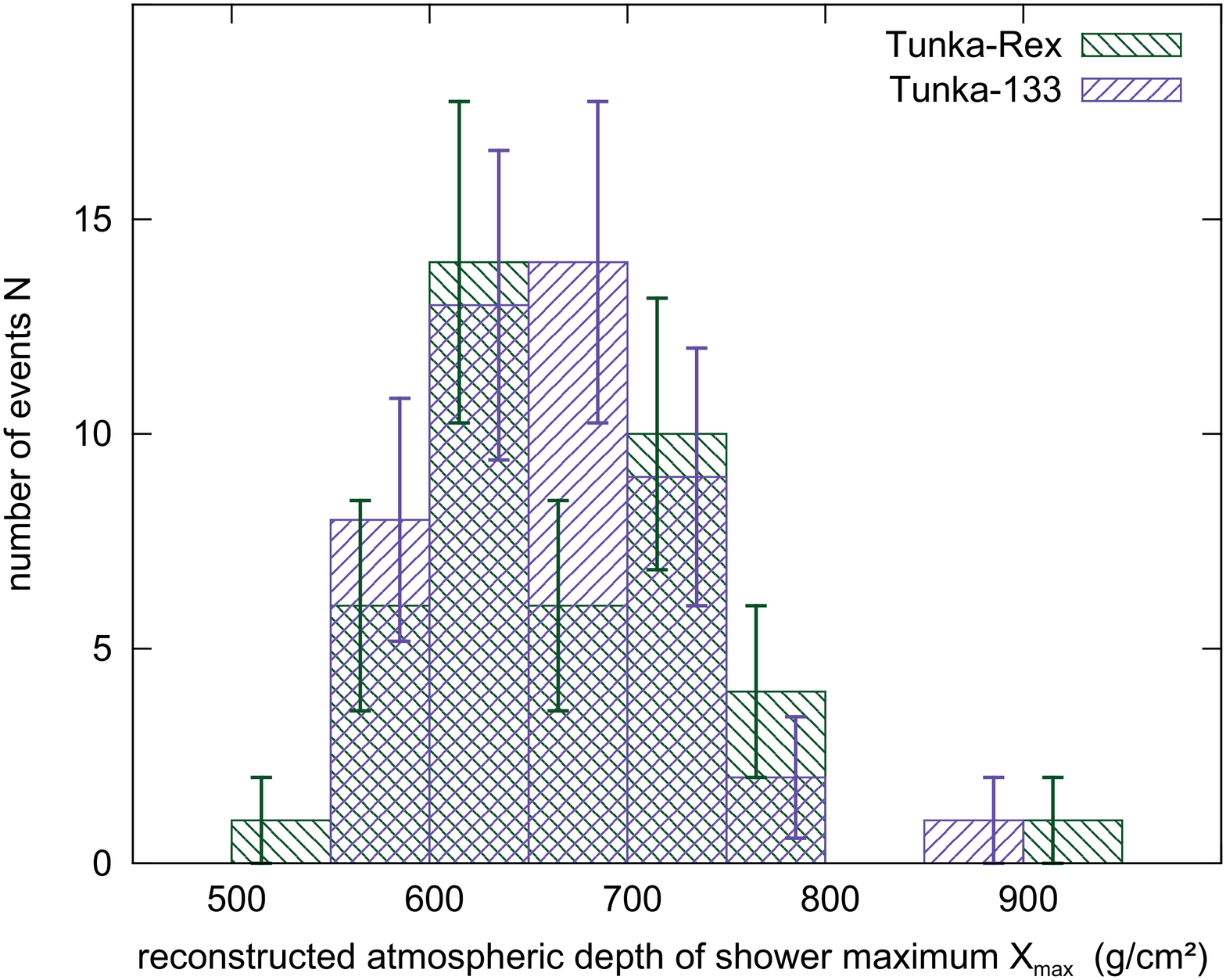}
  \caption{Distributions of energy, $E_\mathrm{pr}$, and atmospheric depth of the shower maximum, $X_{\mathrm{max}}$, of the Tunka-Rex events used in the present analysis. 
  The figure shows the reconstructed values by both Tunka-Rex and Tunka-133. Bars represent the statistical uncertainty calculated as $\sqrt{N}$.}
  \label{fig_distributionHistograms}
\end{figure*}

\begin{figure}
  \centering
  \includegraphics[width=0.6\linewidth]{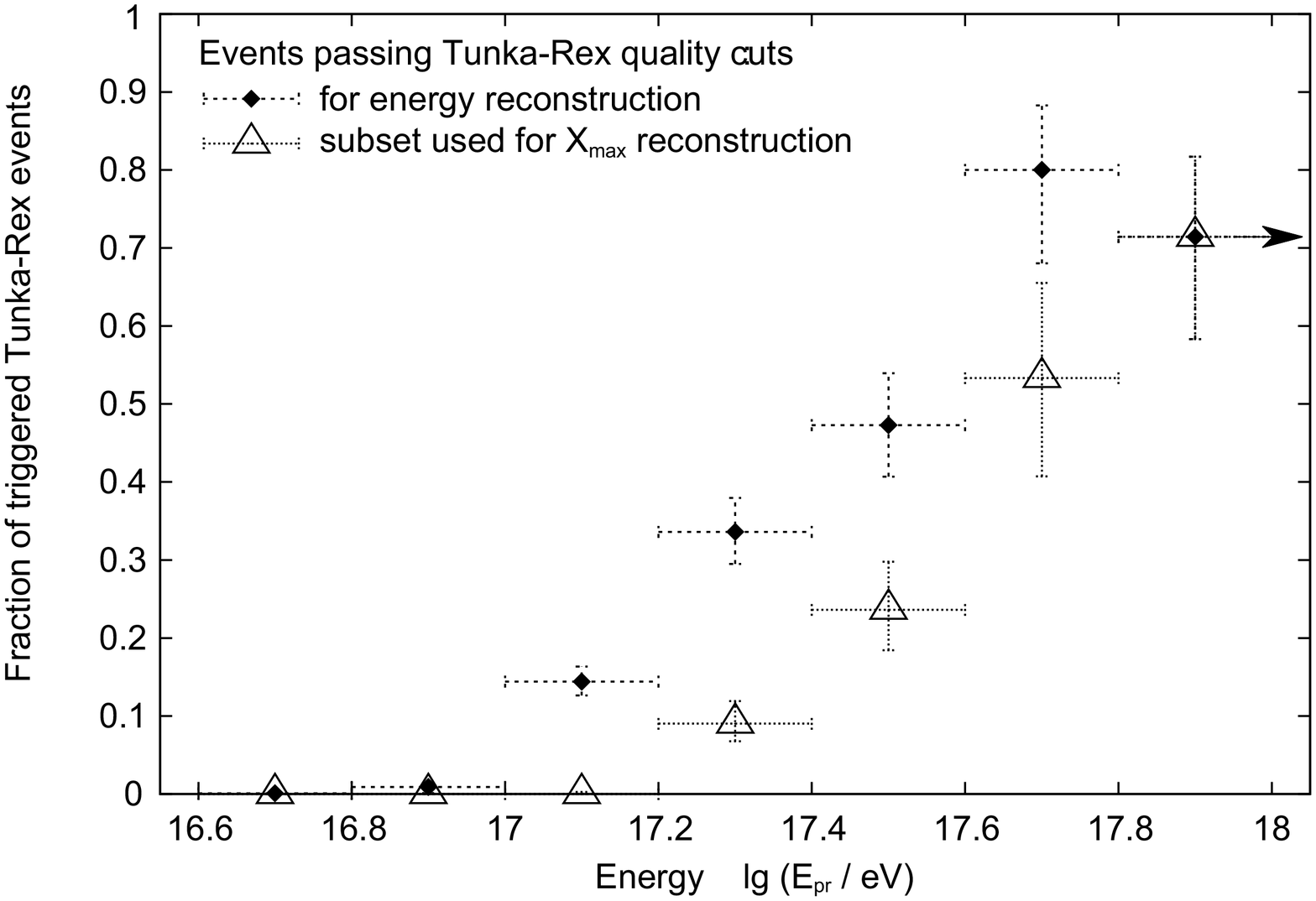}
  \caption{Fraction of events which passes the quality cuts applied for the Tunka-Rex energy and $X_{\mathrm{max}}$ reconstructions, respectively, in comparison with all events triggered by the fully efficient Tunka-133 array in the inner area ($r < 500\,$m).
  }
  \label{fig_FractionOfSelectedEvents}
\end{figure}

An overview of the available event statistics is provided in table \ref{tab_Eventstatistics}, and figure~\ref{fig_distributionHistograms} shows the distribution of the selected events over energy and $X_{\mathrm{max}}$. 
Most of the events used for the energy reconstruction are at energies above $10^{17}\,$eV.
The fraction does not reach $100\,\%$ because some Tunka-133 events have small geomagnetic angles.
Also, there were occasional technical issues in single antenna stations.
For the high-quality events used for the $X_{\mathrm{max}}$ reconstruction, the threshold is slightly higher: above an energy of about $6 \cdot 10^{17}\,$eV, all events used for energy reconstruction pass also the high-quality cuts required for $X_{\mathrm{max}}$ reconstruction (see figure~\ref{fig_FractionOfSelectedEvents}).
Apart from technical issues (e.g., missing or malfunctioning antennas), the fraction shown in figure~\ref{fig_FractionOfSelectedEvents} is the efficiency of Tunka-Rex for zenith angles $\theta \le 50^\circ$, because the reference is the fully efficient Tunka-133 trigger. 
The total number of Tunka-Rex events passing the quality cuts is larger \cite{TunkaRex_NIM_2015}, since in contrast to the air-Cherenkov detection, radio detection is more efficient for inclined air showers. 
Nevertheless, for the present analysis only events with $\theta \le 50^\circ$ have been selected, since Tunka-133 is fully efficient for this zenith angle range. 

\section{Reconstruction of energy and shower maximum}
In the present analysis, the reconstruction of energy and shower maximum is based on the lateral distribution, i.e., the dependence of the radio amplitude on the distance to the shower axis. 
This approach has been used earlier. 
Different experiments have shown that the amplitude at a certain distance is correlated with the primary energy, after correcting for the geomagnetic angle~\cite{Allan1971, FalckeNature2005, Ardouin2009, Glaser_ARENA2012}. 
Moreover, it has been shown that the shape of the lateral distribution depends on the distance to the shower maximum~\cite{2012ApelLOPES_MTD, 2014ApelLOPES_MassComposition, NellesLOFAR_measuredLDF2015}.

Our reconstruction method in brief is:
\begin{enumerate}
 \item We correct the lateral distribution for the azimuthal asymmetry caused by the interference of the geomagnetic and the Askaryan effects, and for the geomagnetic angle~$\alpha$. 
 \item  We fit a Gaussian LDF, the shape of which depends on three parameters: 
 One scale parameter, $\mathcal{E}_{120}$, which is the amplitude at $120\,$m, and two shape parameters $a_1$ and $a_2$, which describe the slope and the width of the Gaussian LDF, respectively. 
 $\mathcal{E}_ {120}$ and $a_1$ are fit to the data points, $a_2$ is fixed using a parametrization depending on energy and zenith angle.
 \item We use the fit result $\mathcal{E}_{120}$ to reconstruct the primary energy $E_\mathrm{pr}$, and the fit result for $a_1$ at $180\,$m axis distance, representing the slope of the lateral distribution, to reconstruct the atmospheric depth of the shower maximum $X_\mathrm{max}$. 
\end{enumerate}

\subsection{Asymmetry correction}
Because of the different polarizations of the radio emission generated by the Askaryan and the geomagnetic effects~\cite{AugerAERApolarization2014, SchellartLOFARpolarization2014, Werner2012}, the radio amplitude on ground depends not only on the geomagnetic angle $\alpha$ and distance to the shower axis $r$, but also on the azimuth angle $\phi_g$ of each antenna relative to the shower axis. 
Fitting the asymmetry for individual events by a two-dimensional lateral distribution function (LDF) introduces additional parameters \cite{NellesLOFAR_measuredLDF2015}. 
To maintain events with low station multiplicity, we have developed an alternative approach described in reference~\cite{KostuninTheory2015}, which is well-suited for sparse arrays like Tunka-Rex.

Our approach makes use of the fact that the asymmetry to first order only depends on the shower geometry, which can be reconstructed independently from the LDF by arrival time measurements.
Based on CoREAS simulations made for the Tunka-Rex setup, the azimuthal asymmetry of the footprint is parametrized as function of the antenna position using a constant typical value of $\varepsilon_\mathrm{asym} = 8.5\,\%$ for the relative strength of the Askaryan effect.\footnote{
Although the size of the asymmetry $\varepsilon_\mathrm{asym}$ depends on zenith angle and distance to the shower axis~\cite{SchellartLOFARpolarization2014}, simulation studies for Tunka-Rex have shown that parameterizing these dependences of $\varepsilon_\mathrm{asym}$ instead of using a constant value does not describe the lateral distribution better. 
Probably this is because Tunka-Rex measurements mostly fall in a limited zenith range. Moreover, $\varepsilon_\mathrm{asym}$ shows little variation in the distance range of $100 - 200\,$m used for energy and $X_{\mathrm{max}}$ reconstruction.} 
Using this parametrization, we correct all measured amplitudes in individual antennas for the azimuthal asymmetry, and obtain a one-dimensional lateral distribution in which the amplitude depends only on the distance to the shower axis. 
Simultaneously, we correct for the relative strength of the geomagnetic Lorentz force, since this depends only on the shower geometry, too, namely on the geomagnetic angle $\alpha$. 
Thus, we use the following equation to obtain the corrected amplitude $\mathcal{E}_\mathrm{cor}(r)$, which is later used to fit a radially symmetric lateral-distribution function:

\begin{equation}
\mathcal{E}_\mathrm{cor}(r) = \mathcal{E}(r, \phi_g) / \sqrt{\varepsilon_\mathrm{asym}^2 + 2 \varepsilon_\mathrm{asym} \cos \phi_g \sin \alpha + \sin^2 \alpha} \,,
\label{eq_assym_correction}
\end{equation}

\noindent
with $\mathcal{E}(r, \phi_g)$ the measured amplitude after noise correction ($\mathcal{E}_\mathrm{true}$ of equation \ref{noise_correction}), $r$ the distance to the shower axis, $\phi_g$ the azimuth angle relative to the shower axis (with $\phi_g = 0$, when a station is in the direction of the geomagnetic Lorentz force), $\alpha$ the geomagnetic angle, $\varepsilon_\mathrm{asym} = 0.085$ the size of the asymmetry, and $\mathcal{E}_\mathrm{cor}(r)$ the amplitude after correction for the asymmetry and the geomagnetic angle.
After this correction of the amplitude measured at each antenna station, the lateral distribution is approximately azimuthally symmetric around the shower axis.
Thus, to first order the amplitude depends only on the distance to the shower axis $r$, and the lateral distribution can be fitted with a one-dimensional function the parameters of which depend on energy and distance to the shower maximum. 

\begin{figure*}[t]
  \centering
  \includegraphics[width=0.99\linewidth]{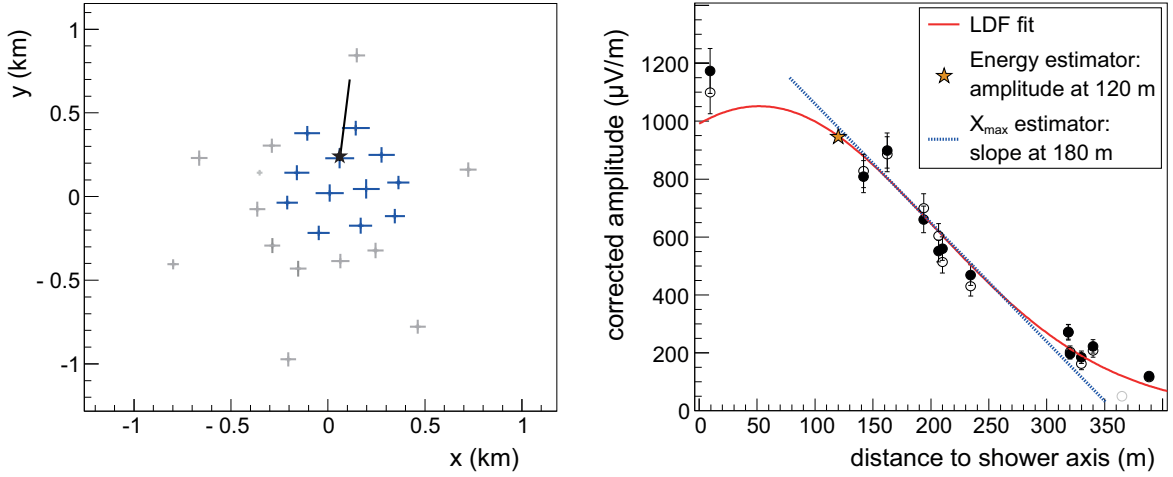}
  \caption{Example event. Left: footprint. Large crosses are triggered stations, and blue crosses are those stations with signal above the signal-to-noise threshold. 
  The small dark-gray cross is a station without measurement for this event. 
  The star and the line indicate the reconstructed shower core and direction.
  Right: Lateral distribution before (open circles) and after (filled circles) asymmetry correction, but with correction for the geomagnetic angle in both cases. The LDF is fitted to the corrected points. The light gray point is a station with signal below the signal-to-noise threshold. }
  \label{fig_exampleEvent}
\end{figure*}

Figure \ref{fig_exampleEvent} shows an example event with many stations, where the asymmetry correction leaves the energy reconstruction almost unchanged, since the LDF fit averages the asymmetry out.
However, the $X_\mathrm{max}$ reconstruction is slightly improved for this event (Tunka-Rex $X_\mathrm{max}$ is $613 \pm 15\,$g/cm\textsuperscript{2} and $621 \pm 15$g/cm\textsuperscript{2} before and after correction, respectively, and the Tunka-133 $X_\mathrm{max}$ is $657\pm28\,$g/cm\textsuperscript{2}).
Especially for events with few stations, the asymmetry correction is important, and can have an effect on the energy reconstruction in the order of $10\,\%$.

\subsection{Lateral distribution function (LDF)}
The amplitude (= electric field strength) of this one-dimensional lateral distribution is proportional to the shower energy, because the radio emission in the shower is coherent. 
Moreover, the shape of the lateral distribution depends mainly on the distance to the shower maximum and has two main features: 
first, a bump caused by Cherenkov-like effects~\cite{AllanICRC1971, Hough1973, Werner2012, NellesLOFAR_CherenkovRing2014}, the existence, size and position of which depend on the distance to the shower maximum; second, an exponential tail~\cite{Allan1971,2010ApelLOPESlateral}, the slope of which depends on the distance to the shower maximum, too. 
This means that even a one-dimensional LDF should contain at least three free parameters: 
one for the energy scale, and two for the shape, namely the width of the bump, and the slope of the exponential fall-off. 
For this purpose we use the following lateral distribution function:

\begin{equation}
\mathcal{E}_\mathrm{cor}(r) = \mathcal{E}_{r_0} \exp\left(a_1(r-r_0) + a_2(r-r_0)^2\right) \,.
\label{general_1dim_ldf}
\end{equation}
with the amplitude $\mathcal{E}_\mathrm{cor}(r)$ after correction for the azimuthal asymmetry and for the geomagnetic angle as function of axis distance $r$, and the fitted scale parameter $\mathcal{E}_{r_0}$, which is the amplitude at the axis distance $r_0$.
The shape parameter $a_1$ determines the slope of the exponential fall-off, and the shape parameter $a_2$ is related to the width of the Cherenkov bump.
The distance parameter $r_0$ is a purely technical parameter.
Using a minimizing chi-square fit, the shape of the function does not depend on the choice of $r_0$, i.e, for different $r_0$ the resulting function is exactly the same, but just described with a different set of values for the parameters $\mathcal{E}_{r_0}$, $a_1$ and $a_2$. 
However, the choice of $r_0$ affects the uncertainties and correlations of the fit parameters, which is why a specific choice of $r_0$ can simplify the reconstruction of $E_\mathrm{pr}$, $X_\mathrm{max}$, and their statistical uncertainties. 
We set this distance parameter $r_0$ to $120\,$m for the reconstruction of the primary energy $E_\mathrm{pr}$ based on $\mathcal{E}_{120}$, and to $180\,$m for the $X_\mathrm{max}$ reconstruction based on $a_1$, as these turned out to be the optimum values in the CoREAS simulation study. 

In the CoREAS simulations made for Tunka-Rex we observe that the scale parameter $\mathcal{E}_{120}$ is strongly correlated with the primary energy $E_\mathrm{pr}$.
The shape parameter $a_1$, i.e., the slope at $180\,$m axis distance is strongly correlated with the distance to the shower maximum. 
The second shape parameter $a_2$ is correlated with $a_1$, and weakly depends on energy and zenith angle.
Thus, fitting $a_2$ as a free parameter would complicate the reconstruction of $X_\mathrm{max}$ significantly. 
For this reason we decided to perform the fit with only two free parameters, and to fix the parameter $a_2$ using a parametrization depending on zenith angle $\theta$ and the primary energy $E_\mathrm{pr}$ (details in appendix~\ref{sec_a2_fixing}).
After fixing $a_2$, the LDF is fit to the data points.
$\mathcal{E}_{120}$ carries the whole sensitivity to the primary energy $E_\mathrm{pr}$, and the remaining free shape parameter $a_1$ carries the whole $X_{\mathrm{max}}$ sensitivity of the LDF. 
Furthermore, reducing the number of free parameters in the LDF from three to two brings the additional advantage that we can use events with only three antenna stations above threshold (about half of the Tunka-Rex events in the present analysis).

\subsection{Primary energy $E_\mathrm{pr}$, and atmospheric depth of shower maximum $X_{\mathrm{max}}$}
Energy $E_\mathrm{pr}$ and $X_\mathrm{max}$ are determined using the equations presented in reference~\cite{KostuninTheory2015}.
While for $X_\mathrm{max}$ we use exactly the same equation as in the reference, i.e., the same values for all parameters in the equation, for $E_\mathrm{pr}$ we use a simplification.
The simplification is to assume that the energy is exactly proportional to the amplitude, as expected for $100\,\%$ coherent emission. 
This means that the exponent $b=0.93$ of reference \cite{KostuninTheory2015} is set to $1$ instead and, thus, is not present in the equation used here.
The reason for applying this simplification is that with the present statistics and quality of the measured data, we have found no evidence for $b \ne 1$. 
Moreover, there is no significant difference in the resulting energy precision whether $b$ is set to $0.93$ or $1$. 
Consequently, at the moment we lack experimental justification for introducing $b$ as an additional parameter and decided for the simpler solution. 
The simplification requires a different proportionality coefficient $\kappa_L$, which has been determined with the same set of CoREAS simulations. 
Summarizing, we use the following equations for the reconstruction of $E_\mathrm{pr}$ and $X_\mathrm{max}$ with Tunka-Rex:

\begin{eqnarray}
E_\mathrm{pr} & = &  \kappa_L \cdot \mathcal{E}_{120} \,,\\
X_\mathrm{max} & = & X_{\mathrm{det}} / \cos\theta - (A + B\log(a_1 + \bar b)) \,,
\label{reconstruction_equation}
\end{eqnarray}
where $X_{\mathrm{det}} = 955$\,g/cm$^2$ is the atmospheric depth of the detector (typical value for the site used in the simulations), and the parameters are determined with CoREAS simulations: 
$\kappa_L = 884\,\mathrm{EeV}/\mathrm{(V/m)}$, $A = -1864\,$\,g/cm$^2$, $B = -566$\,g/cm$^2$, $a_1$ given in units of meters, and $\bar{b} = 0.005$.
As explained above, the technical parameter of the LDF, $r_0$, is set to $120\,$m for the determination of the amplitude at $120\,$m, $\mathcal{E}_{120}$, and to $180\,$m for the determination of the slope parameter $a_1$.

The energy and $X_\mathrm{max}$ reconstructions based on simulations were already presented in reference~\cite{KostuninTheory2015}.
The following section presents the experimental result of Tunka-Rex measurements and the comparison to the Tunka-133 air-Cherenkov measurements.

\section{Results}
\begin{figure*}
  \centering
  \includegraphics[width=0.49\linewidth]{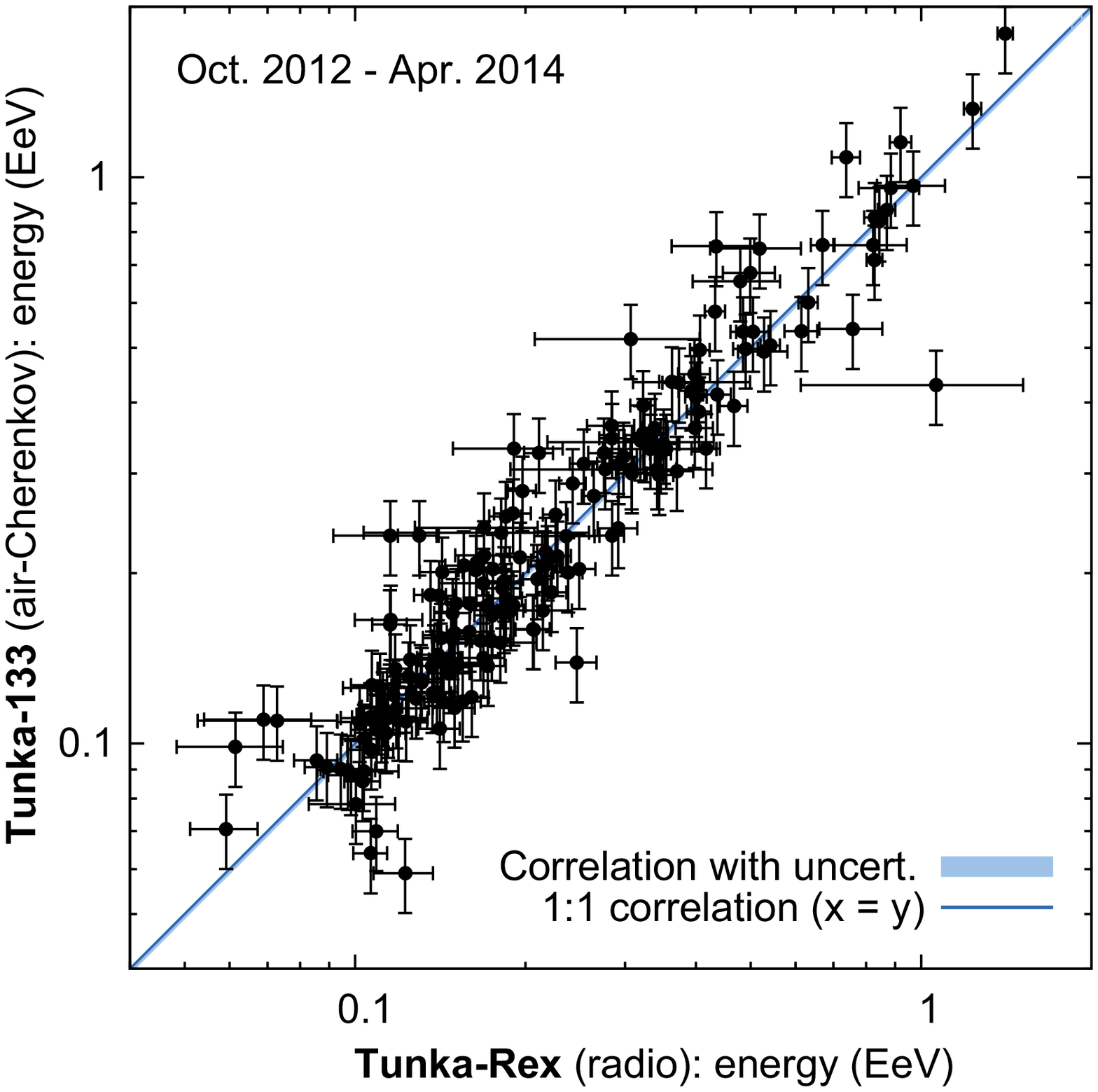}
  \hfill
  \includegraphics[width=0.47\linewidth]{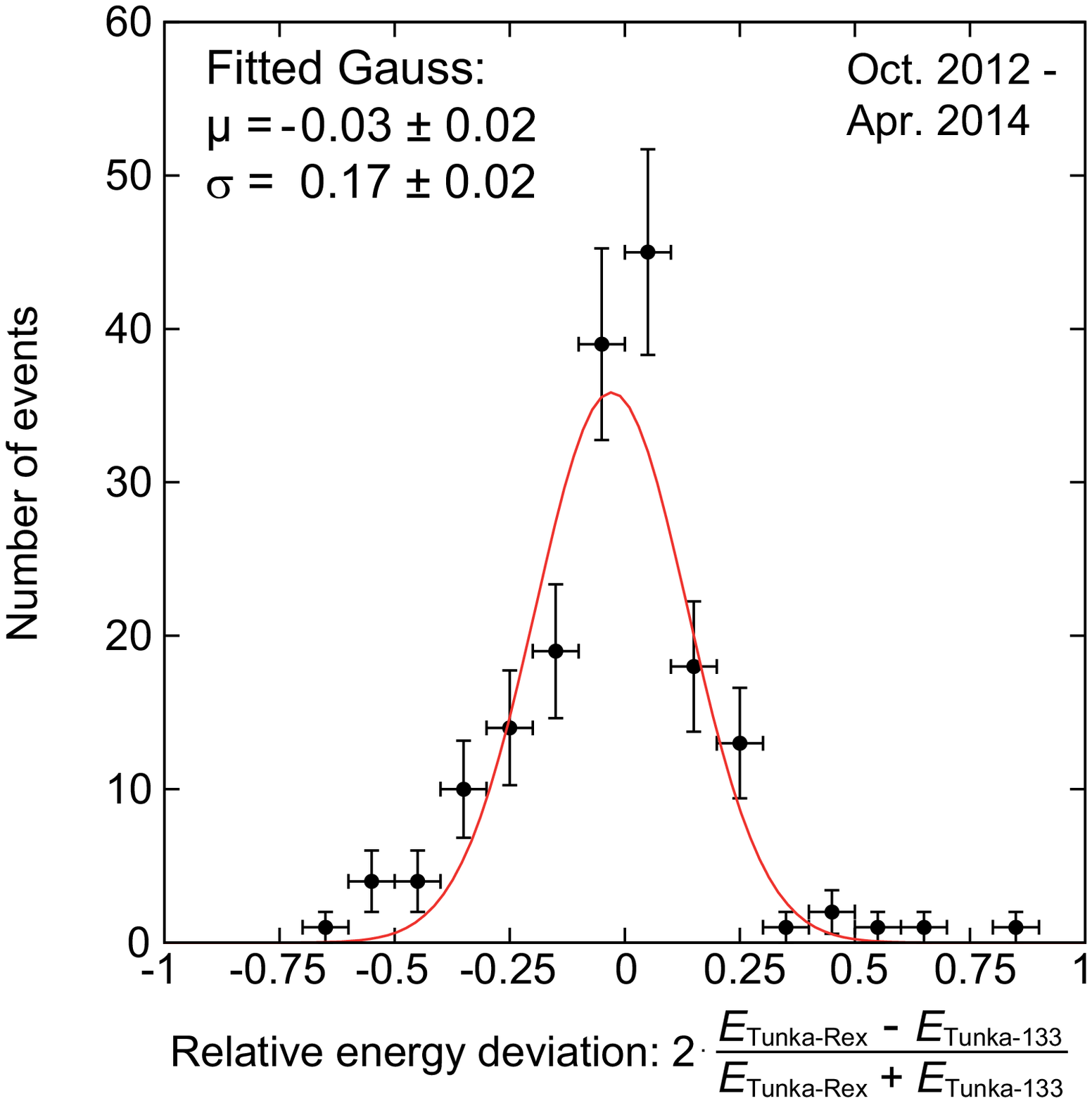}
  \caption{Correlation and relative deviation (= difference divided by average) of the shower energy reconstructed with Tunka-Rex radio and Tunka-133 air-Cherenkov measurements.}
  \label{fig_energyCorrelation}
\end{figure*}

There is a clear correlation between the energy reconstructed from the radio amplitude measured by Tunka-Rex and the energy reconstructed from the air-Cherenkov light measured by Tunka-133 (figure~\ref{fig_energyCorrelation}). 
This indicates that the energy reconstruction works reliably with the presented method. 
When combining both seasons, the average deviation between the Tunka-Rex and Tunka-133 energy values is $(17 \pm 2) \,\%$, which is only slightly larger than the $15\,\%$ energy resolution reported by Tunka-133 \cite{Tunka133_NIM2014}. 

\begin{figure*}
  \centering
  \includegraphics[width=0.49\linewidth]{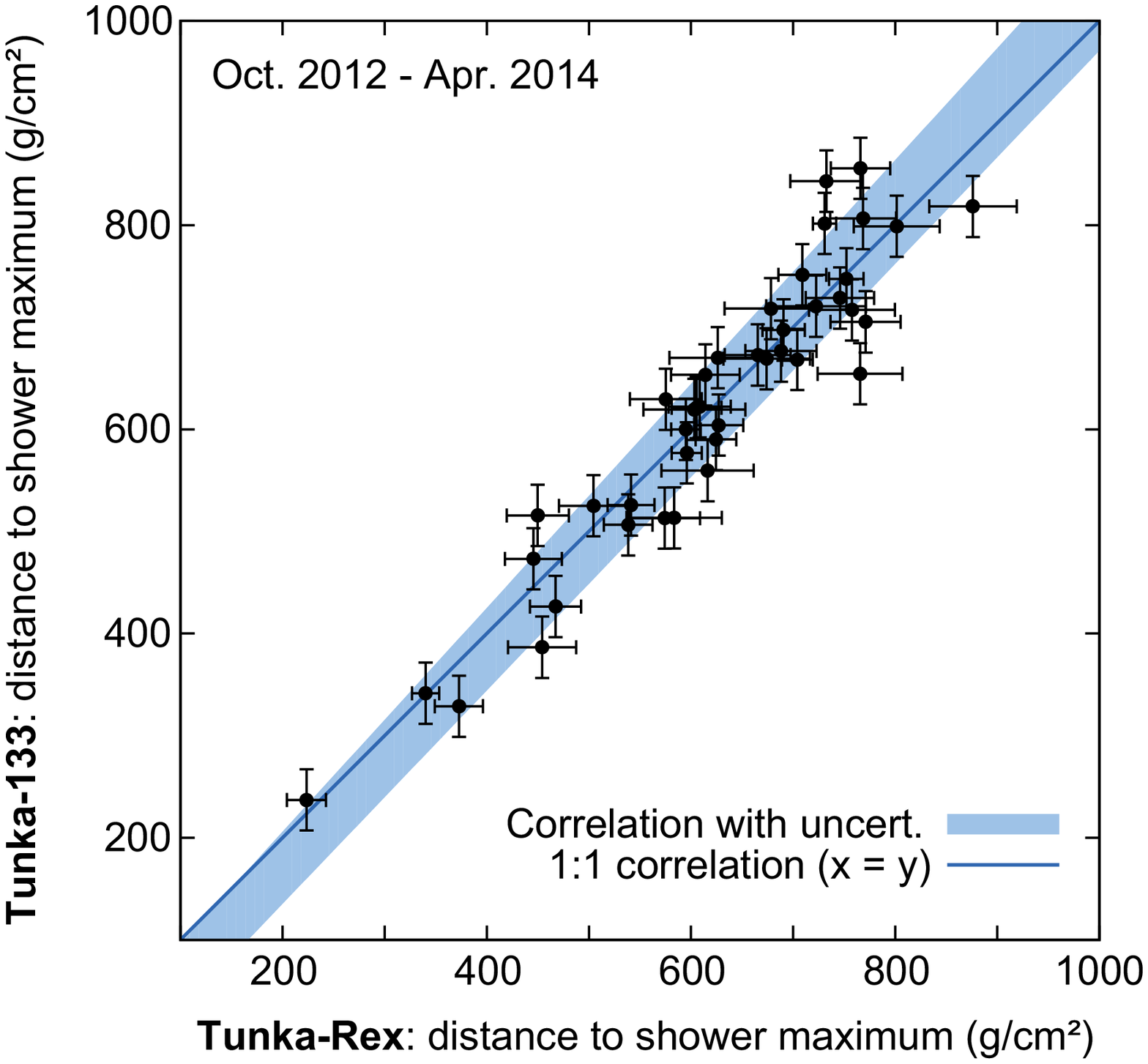}
  \hfill
  \includegraphics[width=0.47\linewidth]{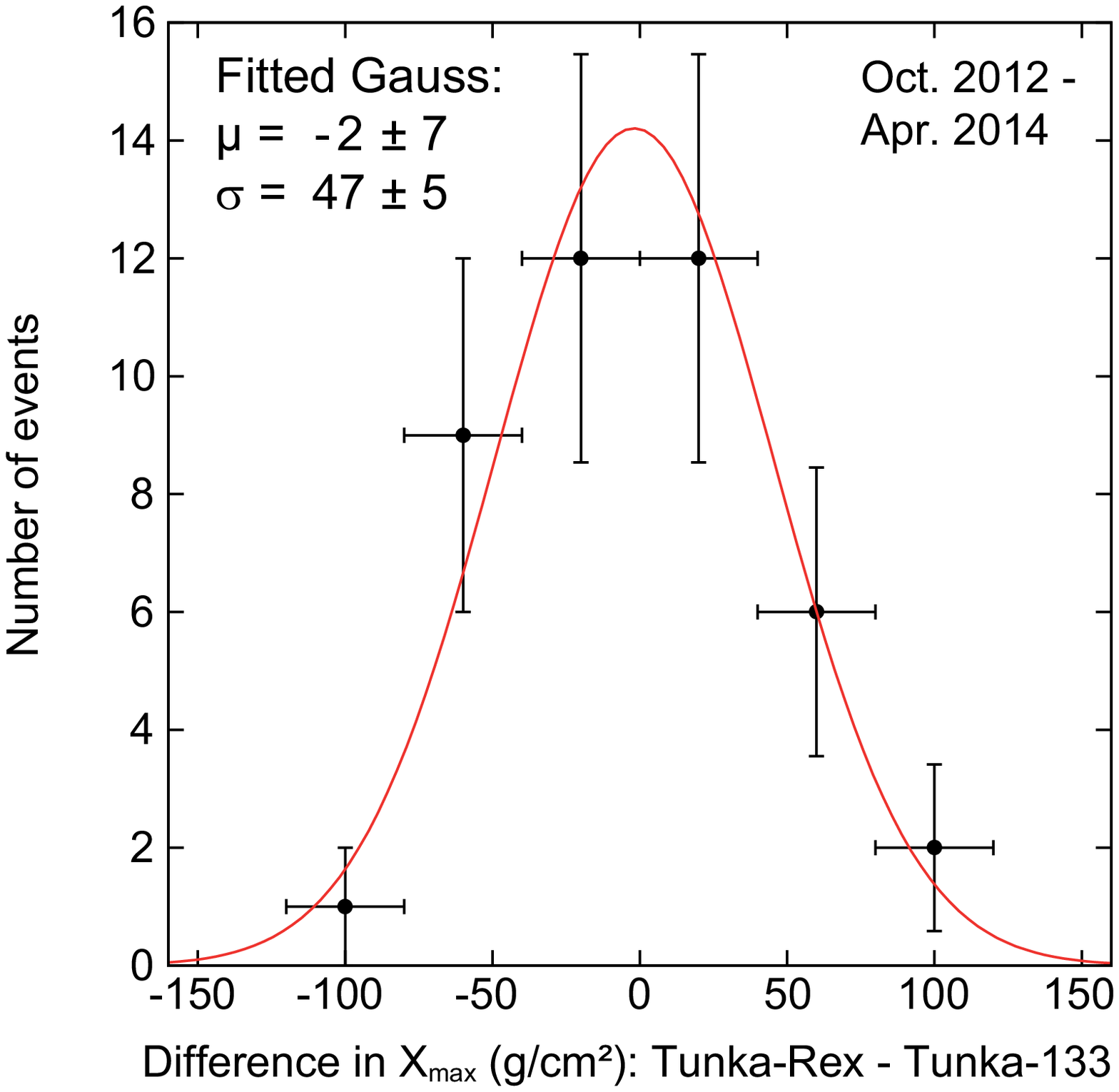}
  \caption{Correlation and difference of the distance to $X_\mathrm{max}$ (atmospheric depth between the shower maximum and the shower core) as reconstructed with Tunka-Rex radio and Tunka-133 air-Cherenkov measurements.}
  \label{fig_XmaxCorrelation}
\end{figure*}

Also for the position of the shower maximum there is a clear correlation between the radio and air-Cherenkov reconstructions (figure~\ref{fig_XmaxCorrelation}). Due to the stricter quality cuts the event statistics is lower than for the energy correlation.
Still, the correlation is significant. 
This again indicates that also the reconstruction of $X_\mathrm{max}$ works reliably for the selected, high-quality radio events. 
Because of the lower statistics, the $X_\mathrm{max}$ reconstruction precision of Tunka-Rex cannot be estimated very accurately. 
The standard deviation of the $X_\mathrm{max}$ difference between both detectors is ($47 \pm 5) \,$g/cm\textsuperscript{2}, which reflects the combination of the unknown Tunka-Rex uncertainty and the Tunka-133 uncertainty of $28\,$g/cm\textsuperscript{2}~\cite{Tunka133_NIM2014}.
This corresponds to a radio-only precision of better than $40\,$g/cm\textsuperscript{2}, when assuming that the resolutions of Tunka-133 and Tunka-Rex add in squares to the total deviation.

For both the energy and $X_\mathrm{max}$ reconstructions the observed correlation is statistically compatible with a 1:1 correlation of Tunka-133 and Tunka-Rex: 
the means of the Gaussians fitted in figures \ref{fig_energyCorrelation} and \ref{fig_XmaxCorrelation} are compatible with 0, where 0 corresponds to no bias (= systematic scale offset). 
This means that the energy and $X_\mathrm{max}$ scales of both detectors agree on an absolute level within the measurement uncertainty. 
This is remarkable, since the Tunka-Rex scale is defined by CoREAS simulations and has not been tuned against Tunka-133.
With already consistent scales there is no need for a true cross-calibration, i.e., en explicit calibration of both detectors against each other.

\section{Discussion}

\subsection{Validity of the results}
The present analysis is the first event-by-event comparison of radio and air-Cherenkov measurements of the same air showers. 
Since both detection methods are sensitive to the energy and the position of the maximum of the electromagnetic shower component, a strong correlation is expected between radio and air-Cherenkov observables.
This is especially noteworthy since the Tunka-Rex reconstruction is based on simulations and not tuned against the Tunka-133 measurements we compared with.
Only details of the reconstruction method have been decided after looking at the data of the first season, e.g., the filter used to remove narrow-band interferences or the exact quality cuts.
Neither of these small decisions made the $X_\mathrm{max}$ and energy correlations appear or vanish. 

As additional cross-check of the validity of the results, we decided to blind the energy and $X_\mathrm{max}$ reconstruction of Tunka-133 for the second season. 
These values were revealed to us only after we froze the reconstruction method, including digital filters and quality cuts, and after we decided to use the simplified equation \ref{reconstruction_equation} for energy reconstruction. 
In appendix~\ref{app_events} the results of both seasons are compared, and found to be compatible within statistical uncertainties. 
Thus, we conclude that our results are not corrupted by any kind of implicit tuning on the data set of the first year. 
Moreover, the quality cuts are general enough to cover the slightly different array configurations in both seasons, i.e., the missing antennas in the first season do not have significant impact on any of the present results. 

Potential biases not excluded by the cross-check concern possible special events. 
For example, events with their shower maximum very close to the detector will have a very steep lateral distribution and a small footprint. 
Thus, they will likely not be detected or will be rejected by the quality cuts. 
Still, since Tunka-133 is fully efficient and has a reliable reconstruction in the full zenith and energy range used in this study, the non-detection of certain events is not assumed to cause any significant bias in the observed correlations.
Furthermore, at high energies less than a third of the events are rejected (cf. figure~\ref{fig_FractionOfSelectedEvents}), and this due to technical reasons (missing antenna stations) or due to small geomagnetic angles leading to small radio amplitudes. 
Thus, the potential problem of a bias introduced by the rejection of any kind of special event can concern only a small fraction of events. 
Nevertheless, this has to be studied in more detail when larger statistics will be available. 

Consequently, the cross-check of both seasons cannot totally exclude all imaginable systematic biases implicit in the experimental setup or in the simulations simulations, but it definitely confirms that the observed correlations are real.

\subsection{Precision and accuracy}
The spread of the deviation between the Tunka-Rex and Tunka-133 reconstruction values for the same events can be used to estimate the reconstruction precision of Tunka-Rex. 
Assuming independent Gaussian uncertainties of both reconstructions, the standard deviation of the difference or ratio between Tunka-Rex and Tunka-133 is the quadratic sum of the absolute or relative uncertainties, receptively.
If Tunka-Rex would feature the same precision as Tunka-133, thus, this standard deviation should be a factor of $\sqrt{2}$ larger than the Tunka-133 precision alone, i.e.: $\sqrt{2} \cdot 15\,\% = 21\,\%$ for the energy, which is slightly larger than the observed standard deviation of $(17 \pm 2)\,\%$.
Consequently, the energy precision of Tunka-Rex seems to be at least equal to that of Tunka-133. 

For $X_{\mathrm{max}}$, we would expect a standard deviation of the difference of $\sqrt{2} \cdot 28\,$g/cm$^2 = 40\,$g/cm$^2$ if Tunka-Rex had the same precision as Tunka-133 ($28\,$g/cm$^2$), but observe a standard deviation of the difference of $(47 \pm 5)\,$g/cm$^2$.
The derived Tunka-Rex precision is slightly better than $40\,$g/cm$^2$, as expected from CoREAS simulations when including realistic background \cite{KostuninTheory2015}. 
Since the average spacing of the Tunka-133 photomultipliers is half of the average antenna spacing, this result does not necessarily imply that the radio $X_{\mathrm{max}}$ precision is intrinsically worse than the air-Cherenkov precision.
Instead, that later studies will show how the Tunka-133 and Tunka-Rex resolutions compare for equal detector spacing, and by how much the Tunka-Rex precision will improve due to additionally deployed antennas. 

Using CoREAS simulations we searched for systematic uncertainties potentially degrading the resolution further, in particular biases of the $X_{\mathrm{max}}$ reconstruction over energy or zenith angle, but we did not find any significant effect. 
The finally achievable $X_{\mathrm{max}}$ uncertainty of the radio technique might become equal to the one of the air-Cherenkov technique, when events of even higher quality are used. 
The spread between the Tunka-133 and Tunka-Rex $X_{\mathrm{max}}$ values approximately corresponds to the cut placed on the estimator $\sigma(X_\mathrm{max}^\mathrm{radio})$ for the statistical uncertainty of Tunka-Rex. 
This means that this estimator is reliable and can be used to set stronger quality cuts when higher statistics are available.

In addition to the precision also the total accuracy is of importance. 
This is dominated by the $18\,\%$ scale uncertainty of the amplitude calibration \cite{TunkaRex_NIM_2015}, but at least three further systematic scale uncertainties have to be considered: 

First, if the fraction of the total primary energy going in the electromagnetic shower component was simulated wrongly in CORSIKA, the energy scale would be wrong by roughly the same value.
However, the electromagnetic shower component is relatively well understood, and as estimated by other experiments~\cite{AugerNIM2015}, this uncertainty on the \lq invisible\rq~energy is only a few percent. 
Moreover, we assume that the CoREAS radio extension of CORSIKA simulates the absolute radio amplitude emitted by the electromagnetic shower component correctly, i.e., with errors which are small against the calibration scale uncertainty of $18\,\%$, which is supported by recent experimental tests \cite{TunkaRex_NIM_2015, 2015ApelLOPES_improvedCalibration}.
Still, the fraction of energy going in the electromagnetic shower component depends on the mass of the primary particle.
Thus, the parameters in the method slightly depend on the assumed mass composition of the primary particles, in particular also the parameter $\kappa_L$ determining the energy scale.
For the present analysis, we determined $\kappa_L$ as average of proton and iron simulations with the hadronic interaction model QGSJET-II.04~\cite{OstapchenkoQGSjetII2006}. 
If we had taken a pure proton or a pure iron composition to determine $\kappa_L$ of equation \ref{reconstruction_equation}, the reconstructed energy would be $4\,\%$ smaller or larger, respectively.

Second, there is some freedom in the equation used for energy reconstruction. 
For example, if we had used exactly the energy equation of reference \cite{KostuninTheory2015} instead of equation \ref{reconstruction_equation}, then the reconstructed primary energy would be about $10\,\%$ lower, although both equations have been tuned against the same set of CoREAS simulations.
This difference is well within the $18\,\%$ scale uncertainty of the amplitude calibration, but indicates that a more detailed study of the reconstruction method will become necessary once a more accurate calibration becomes available.

Third, using CoREAS simulations we found a systematic bias of the energy reconstruction depending on the distance to the shower maximum. 
The mean effect is smaller than the difference between proton- and iron-initiated showers, and thus negligible. 
However, for very close or far distances to the shower maximum ($\lesssim 350\,$g/cm$^2$ or $\gtrsim 900\,$g/cm$^2$) the bias becomes larger than $10\,\%$, and will have to be corrected for in such rare events. 
Consequently, all known systematic uncertainties are small compared to the scale uncertainty of the amplitude calibration of $18\,\%$. 
Thus, the total accuracy of the Tunka-Rex energy scale is approximately $20\,\%$. 

The Tunka-Rex energy scale defined by CoREAS simulations agrees within the uncertainties to the Tunka-133 scale.
This means that the measurements of Tunka-Rex and Tunka-133 are not only correlated, but in addition agree on an absolute level, which is another confirmation, that CoREAS seems to predict the radio signal correctly.

\subsection{Comparison with other experiments}
The $15\,\%$ energy precision of Tunka-Rex is at least as good as that of other radio arrays, like AERA~\cite{AugerAERAenergy2015}, LOPES~\cite{2014ApelLOPES_MassComposition}, and CODALEMA~\cite{RebaiCODALEMAenergy2012}, which all reported energy precisions around $20\,\%$.  
The Tunka-Rex scale uncertainty of $20\,\%$ roughly corresponds to that of AERA~\cite{AugerAERAenergy2015} and LOPES~\cite{2015ApelLOPES_improvedCalibration}, which compared the radio measurements to air-fluorescence and particle measurements on ground. 
LOFAR as well features an absolute calibration of the radio amplitude with similar accuracy~\cite{NellesLOFARcalibration2015}, but experimental checks of the energy accuracy are limited by the poor energy resolution of the LORA particle detector array \cite{NellesLOFAR_measuredLDF2015}. 
For $X_{\mathrm{max}}$ reconstruction, the resolution of LOPES has been significantly worse, using a simpler method in a more radio-loud environment than Tunka-Rex. 
The much denser array LOFAR is located in a more radio-quiet environment similar to the situation of Tunka-Rex. 
Its $X_{\mathrm{max}}$ reconstruction is also based on CoREAS simulations using two different methods.
With an $X_{\mathrm{max}}$ reconstruction based on a fitted LDF (like in our approach), the precision of $X_{\mathrm{max}}$ at LOFAR is only slightly better than our resolution \cite{NellesLOFAR_measuredLDF2015}. 
However, with a more computing-intensive method using many simulations for each individual event \cite{BuitinkLOFAR_Xmax2014}, the precision by LOFAR is at least twice as good as ours. 
This indicates that in addition to the planned deployment of additional antennas, also further improvements in the reconstruction methods could increase the $X_{\mathrm{max}}$ resolution of Tunka-Rex.

The most important scientific value of the present analysis is the experimental comparision of radio measurements against another, established technique. 
For the first time, energy and $X_{\mathrm{max}}$ reconstructions based on absolutely-calibrated radio measurements have been compared to air-Cherenkov measurements.
While the principle sensitivity of the radio signal on the longitudinal shower development was already demonstrated experimentally by comparing LOPES lateral distributions to measurements of the KASCADE muon-tracking detector~\cite{2012ApelLOPES_MTD}, KASCADE did not feature direct $X_{\mathrm{max}}$ measurements. 
Thus, with the present comparison of Tunka-Rex and Tunka-133, for the first time the radio reconstruction of $X_{\mathrm{max}}$ is directly cross-checked with an independent $X_{\mathrm{max}}$ measurement.
This also gives more confidence in the results of other radio arrays whose reconstruction procedures are developed with the same CORSIKA + CoREAS Monte Carlo codes.

\section{Outlook}
The presently achieved accuracy for energy and $X_{\mathrm{max}}$ reconstruction by Tunka-Rex is not yet at its principle limits. 
In particular for the $X_{\mathrm{max}}$ reconstruction, there is room for improvement.
As shown by LOFAR \cite{BuitinkLOFAR_Xmax2014}, the precision can be as good as $20\,$g/cm\textsuperscript{2}, when using a denser radio array and a simulation-driven method. 
Thus, we plan to deploy additional antennas and to further improve the reconstruction method. 
With an increased antenna density, the average event will contain more antenna stations with signal, which allows fitting of lateral distributions with more free parameters.
For example, the weak energy and $X_{\mathrm{max}}$ sensitivity of the LDF parameter $a_2$ could be exploited when more antenna stations contribute to the fit.

Compared to this study the antenna density will be tripled. 
At each of the 19 Tunka clusters in the inner area, already now a second antenna station is deployed, and a third one will be deployed in summer 2016.
This enables an experimental test how the antenna density and the core resolution affect the $X_{\mathrm{max}}$ resolution. 
Moreover, additional methods for $X_{\mathrm{max}}$ reconstruction based on other quantities of the radio signal can be used to improve the total accuracy. 
The shape of the wavefront~\cite{2014ApelLOPES_wavefront} can be reconstructed by arrival time measurements, and the slope of the frequency spectrum~\cite{Grebe_ARENA2012} by sampling the pulse shape. 
Future studies will show by how much these methods can increase the total $X_{\mathrm{max}}$ accuracy under practical conditions.

The statistics of Tunka-Rex will be dramatically increased by joint measurements with the newly deployed particle-detector array Tunka-Grande~\cite{TAIGA_2014}, which provides a day-time trigger for Tunka-Rex. 
Tunka-Grande consists of 19 stations, one at each inner cluster, where each station features scintillator detectors on ground and under ground, for measurement of secondary air-shower electrons and muons, respectively. 
While unimportant for the present analysis focused on the feasibility of the reconstruction methods, for reconstruction of the cosmic-ray composition possible systematic uncertainties like selection biases are relevant.
Using the additional statistics and independent measurements provided by Tunka-Grande such systematic uncertainties can be checked experimentally.  
Moreover, the additional muon measurements can enhance the total accuracy for the mass composition, since the electron-muon ratio provides complementary mass information to $X_{\mathrm{max}}$~\cite{KampertUnger2012}.
In addition, the Tunka-Grande measurement accuracy will be enhanced by a cross-calibration on hybrid measurements with Tunka-Rex.

Finally, the absolute energy measurement of air-showers with a radio array opens prospects to cross-calibrate the energy scales of different air-shower experiments via radio measurements. 
This aspect is worth to be investigated more deeply, in particular since antenna arrays can be a very economic add-on to existing detectors as shown by Tunka-Rex.

\section{Conclusion}
We have compared energy and $X_{\mathrm{max}}$ reconstructed from Tunka-Rex and Tunka-133 measurements of the same air showers in the energy range $E_\mathrm{pr} \gtrsim 10^{17}\,$eV. 
The reconstruction methods have been developed and tuned using simulations with CORSIKA and its radio extension CoREAS. 
For both parameters we find a strong correlation between Tunka-Rex and Tunka-133 consistently in two seasons of data taking. 
This confirms experimentally that the radio measurements are indeed sensitive to energy and shower maximum. 
For $X_{\mathrm{max}}$ this is the first direct confirmation based on a cross-check with a different, established experimental technique, namely air-Cherenkov measurements. 

The Tunka-Rex energy precision seems to be at least as good as the published Tunka-133 resolution of $15\,\%$.
The total scale uncertainty of Tunka-Rex is dominated by the uncertainty of the amplitude calibration, and in total is in the order of $20\,\%$. 
This is comparable to the scale uncertainty of particle detector arrays, like KASCADE-Grande~\cite{2012ApelKGenergySpectrum}.
Since the calibration uncertainty dominates the total scale uncertainty of Tunka-Rex, any improvement in the calibration accuracy will directly propagate to the energy-scale accuracy.
Consequently, further efforts on the calibration will be necessary to increase the accuracy to the same level as the currently leading fluorescence and air-Cherenkov techniques. 
At the Pierre Auger Observatory, the scale accuracy of fluorescence measurements is $14\,\%$~\cite{AugerNIM2015}.
This now is in reasonable reach for future radio measurements, which unlike fluorescence measurements are available around-the-clock. 
Hence, radio measurements can be used to determine the absolute energy scale of air-shower measurements.

The $X_{\mathrm{max}}$ precision of Tunka-Rex is roughly $40\,$g/cm\textsuperscript{2}, and can be slightly increased by setting stricter quality cuts at the cost of statistics.
This resolution is sufficient to statistically distinguish light from heavy primary particles. 
The resolution is worse than the one currently achieved by non-imaging air-Cherenkov measurements ($28\,$g/cm\textsuperscript{2} for Tunka-133~\cite{Tunka133_NIM2014}) and by air-fluorescence measurements ($20\,$g/cm\textsuperscript{2} for the Pierre Auger Observatory~\cite{AugerNIM2015}).
Nevertheless, the resolution of Tunka-Rex is not yet at its limit and can likely be increased by further improvements of the reconstruction method, and by deploying additional antennas. 

For astrophysical applications the total accuracy for the reconstruction of the mass composition as a function of energy counts, e.g., to better study the transition from Galactic to yet-unknown extra-galactic cosmic-ray sources assumed in the energy range of Tunka-Rex~\cite{2013ApelKG_LightAnkle}. 
Tunka-Rex will provide additional statistics exactly in the energy range where current Tunka-133 analyses are limited by statistics. 
Moreover, hybrid measurements of air-showers by Tunka-Rex and Tunka-Grande started in autumn 2015. 
These hybrid measurements can be used to enhance the total accuracy for the mass composition by combining radio and muon measurements. 

\acknowledgments
Tunka-Rex has been funded by the German Helmholtz association and the Russian Foundation for Basic Research (grant HRJRG-303). 
Moreover, this work was supported by the Helmholtz Alliance for Astroparticle Physics (HAP), and by the Russian grant RSF 15-12-20022.

\appendix
\section{Parametrization of LDF parameter $a_2$}
\label{sec_a2_fixing}

In the Gaussian lateral distribution function (LDF, equation \ref{general_1dim_ldf}), the parameter $a_2$ defining the bump has been set by the following parameterization determined with CoREAS simulations made for the situation of Tunka-Rex \cite{KostuninTheory2015}:

\begin{eqnarray}
a_2(\theta, E_\mathrm{pr}) & = & a_{21}(E_\mathrm{pr}) - a_{22}(E_\mathrm{pr})\cos\theta\,,\\
a_{21} & = & 0.19\cdot 10^{-5}\,\frac{1}{\mathrm{m}^2} - 1.63\cdot 10^{-5}\,\frac{E_\mathrm{pr}}{\mathrm{m}^2 \cdot \mathrm{EeV}} \,,\\ \nonumber
a_{22} & = & -3.45\cdot 10^{-5}\,\frac{1}{\mathrm{m}^2} + 2.44\cdot 10^{-5}\,\frac{E_\mathrm{pr}}{\mathrm{m}^2 \cdot \mathrm{EeV}} \,. \nonumber
\label{sigma_dependence}
\end{eqnarray}

Since the $a_2$ parameter is a function of the primary energy $E_\mathrm{pr}$, the shape of the Gaussian LDF used to reconstruct the primary energy depends implicitly on the energy itself. 
This reciprocal dependence on $E_\mathrm{pr}$ can either be solved by an iterative approach, or by using a pre-estimate for $E_\mathrm{pr}$. 
We decided for the latter solution, using the energy estimator based on a simpler exponential LDF presented in reference \cite{TunkaRex_NIM_2015}, since its precision is only slightly worse than than in our more advanced approach here.

\begin{figure*}
  \centering
  \includegraphics[width=0.47\linewidth]{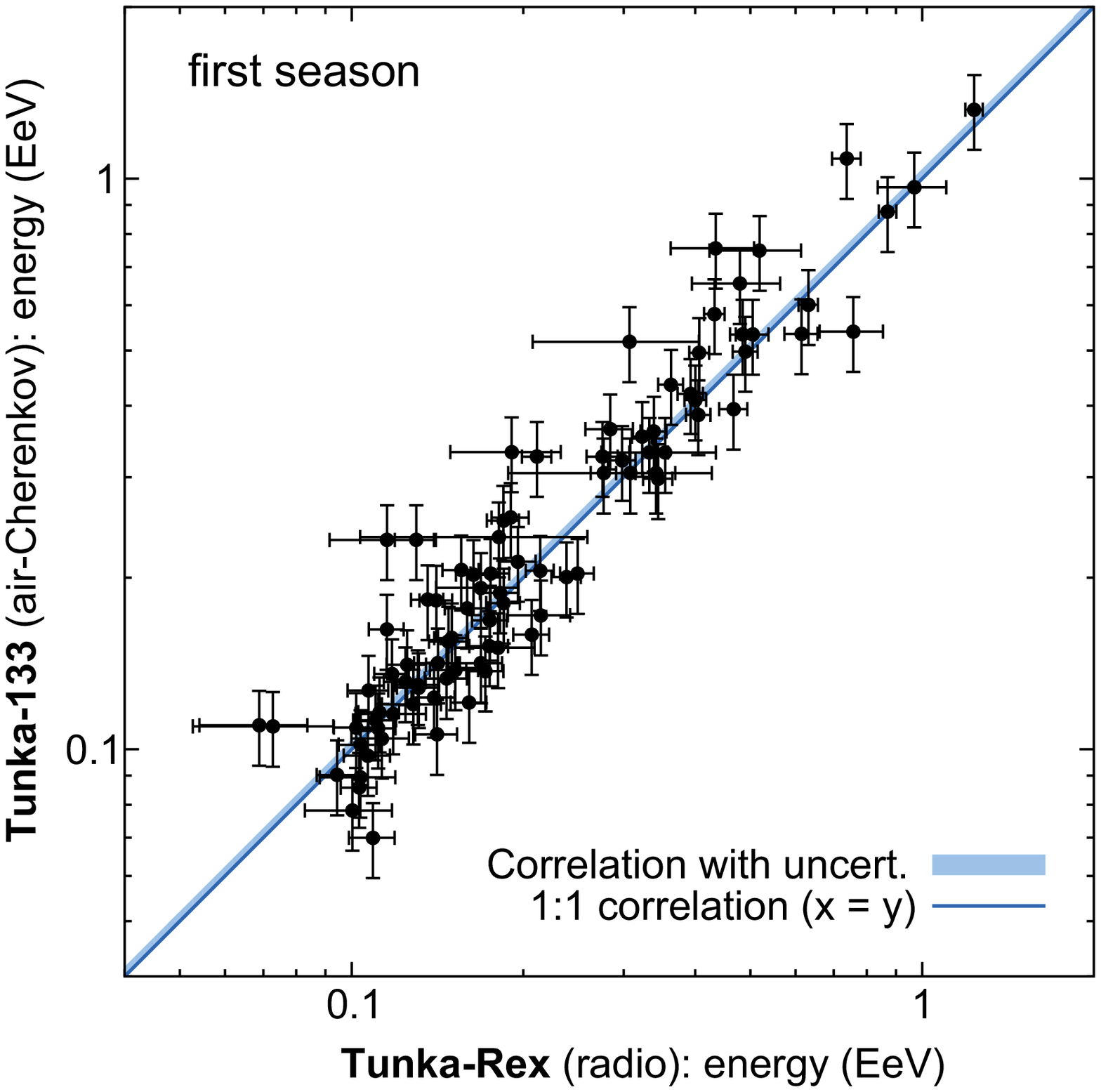}
  \hfill
  \includegraphics[width=0.47\linewidth]{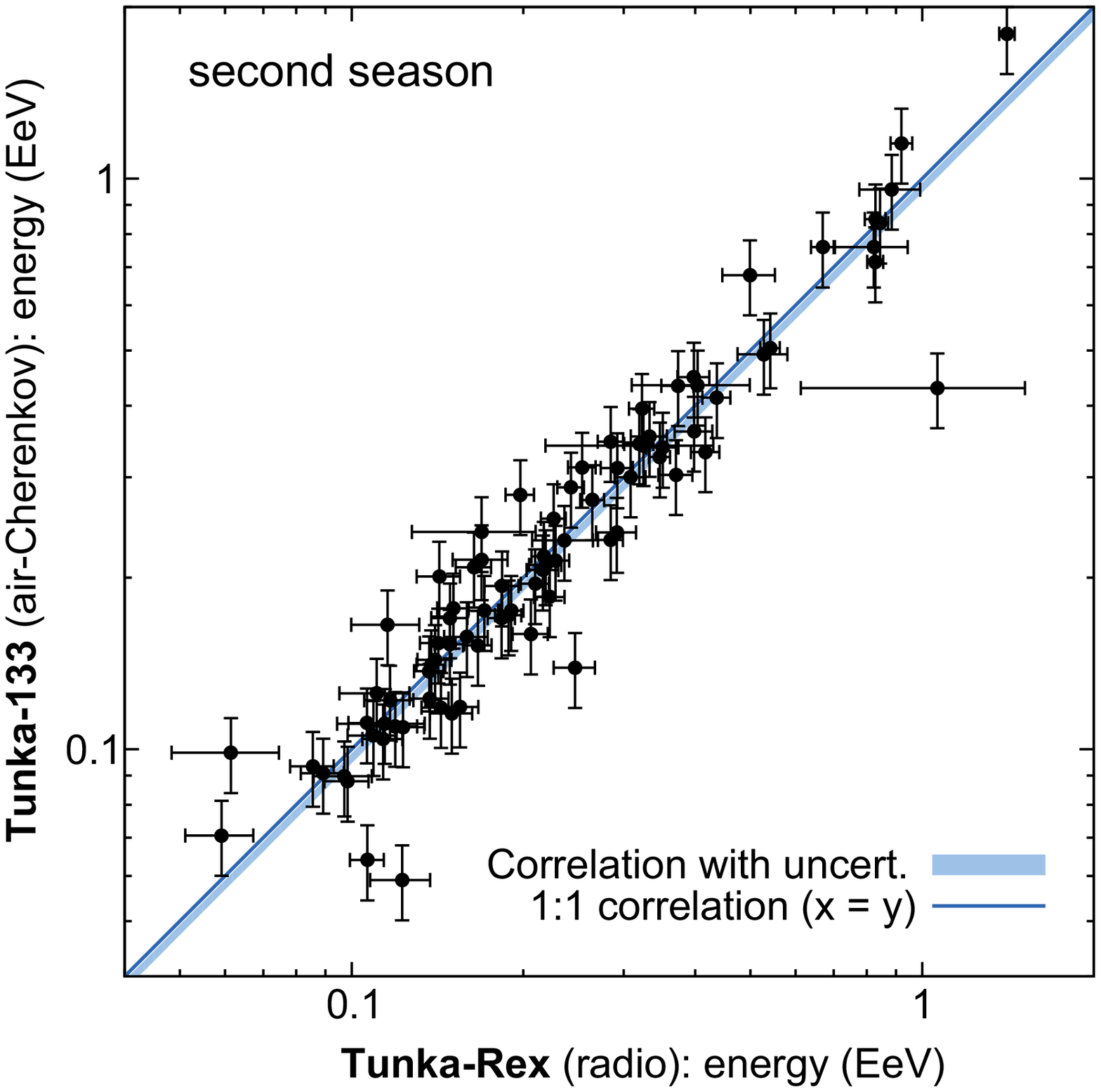}
  \includegraphics[width=0.49\linewidth]{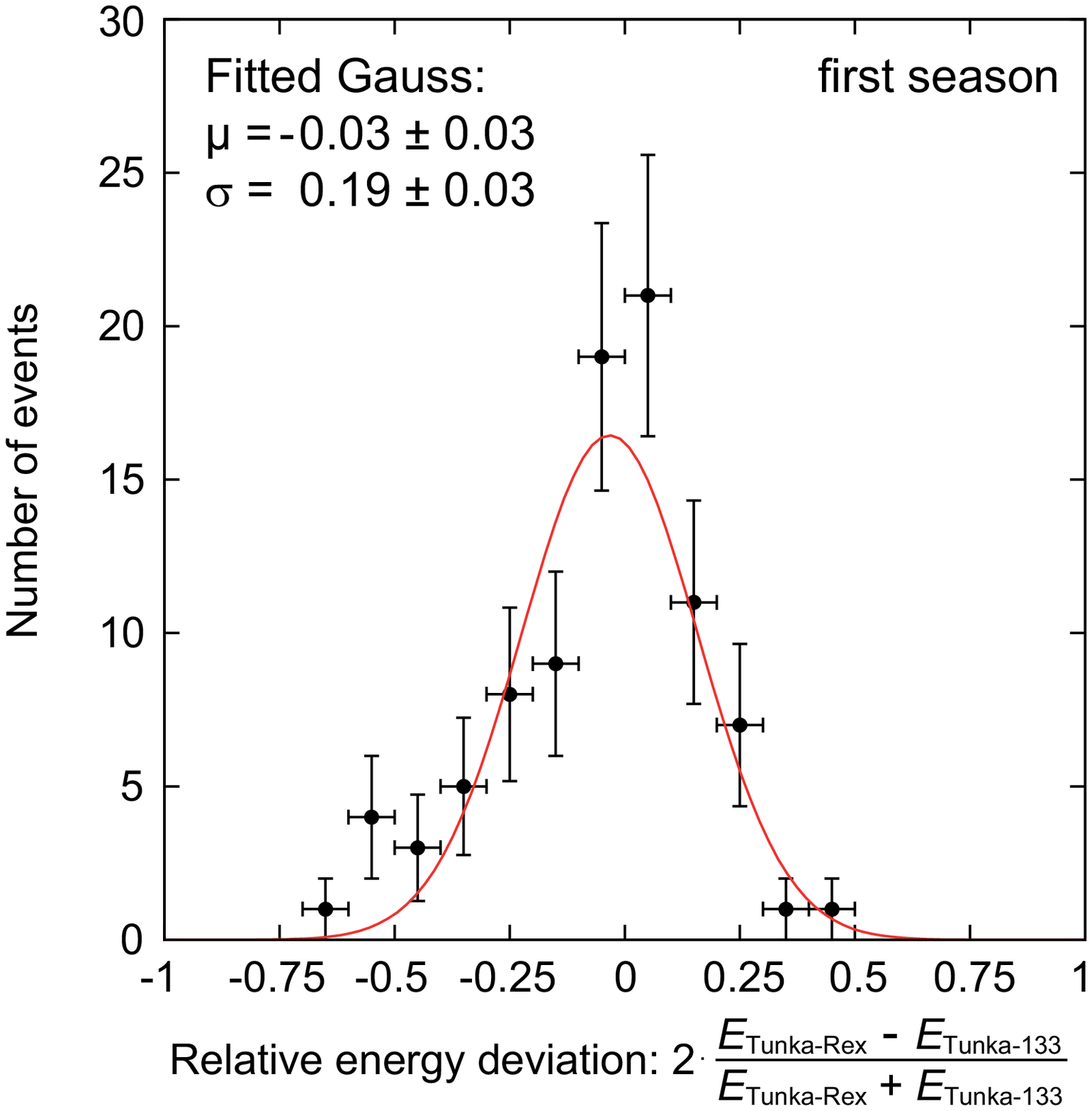}
  \hfill
  \includegraphics[width=0.49\linewidth]{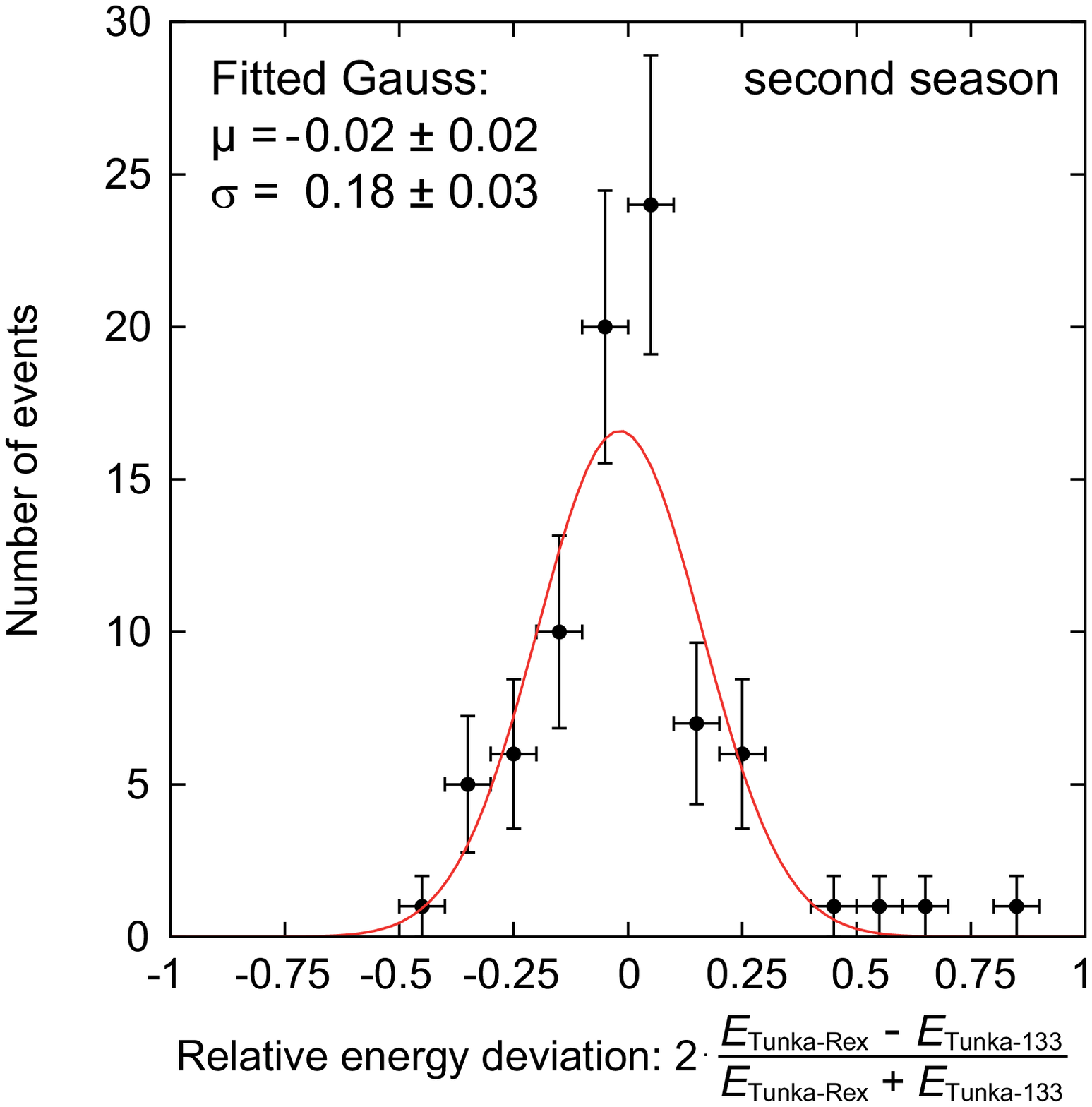}
  \caption{Comparison of the first and second season: reconstructed energy (cf.~figure~\ref{fig_energyCorrelation}).}
  \label{fig_energyCorrelationCompareSeasons}
\end{figure*}

\begin{figure*}
  \centering
  \includegraphics[width=0.49\linewidth]{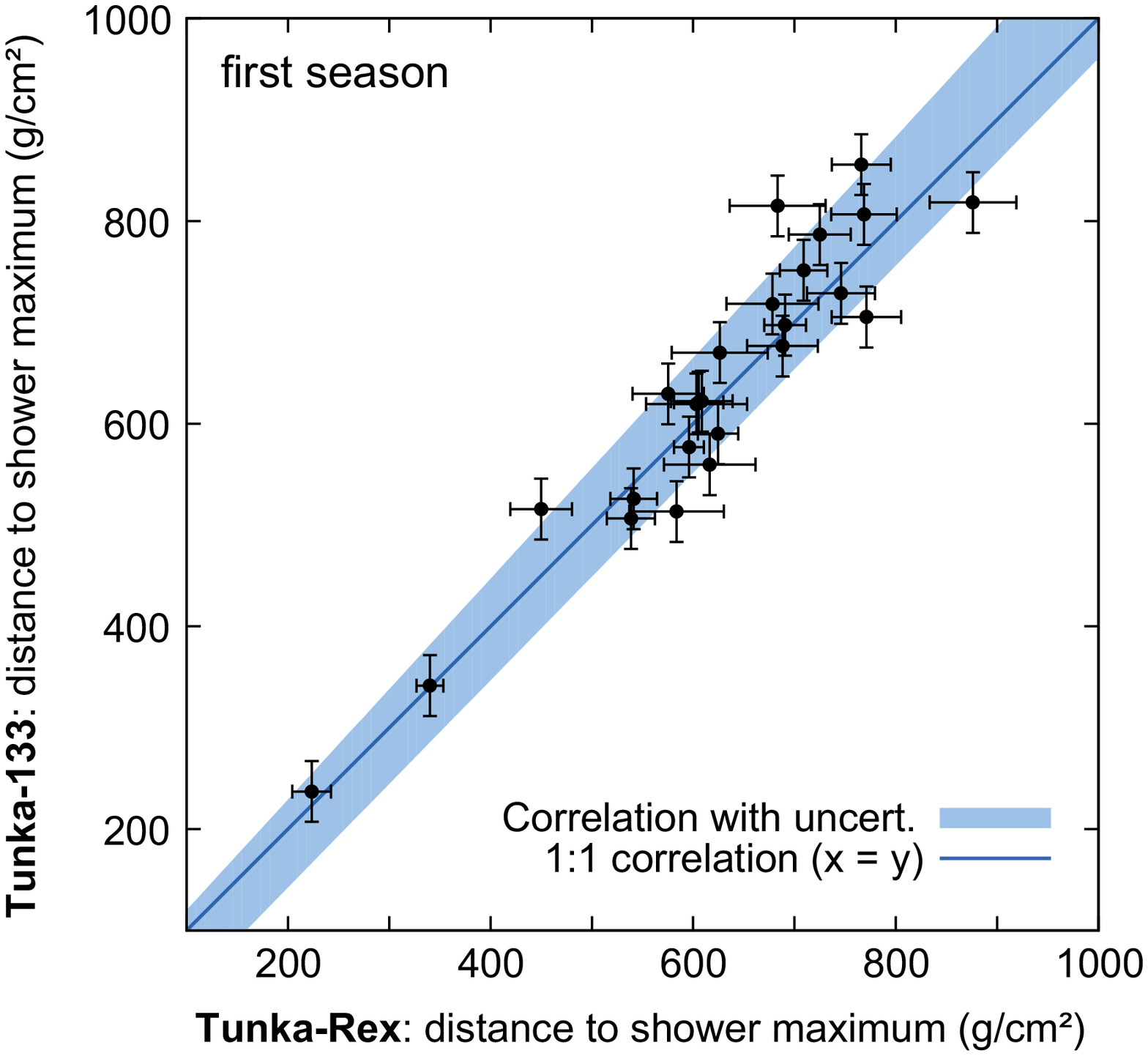}
  \hfill
  \includegraphics[width=0.49\linewidth]{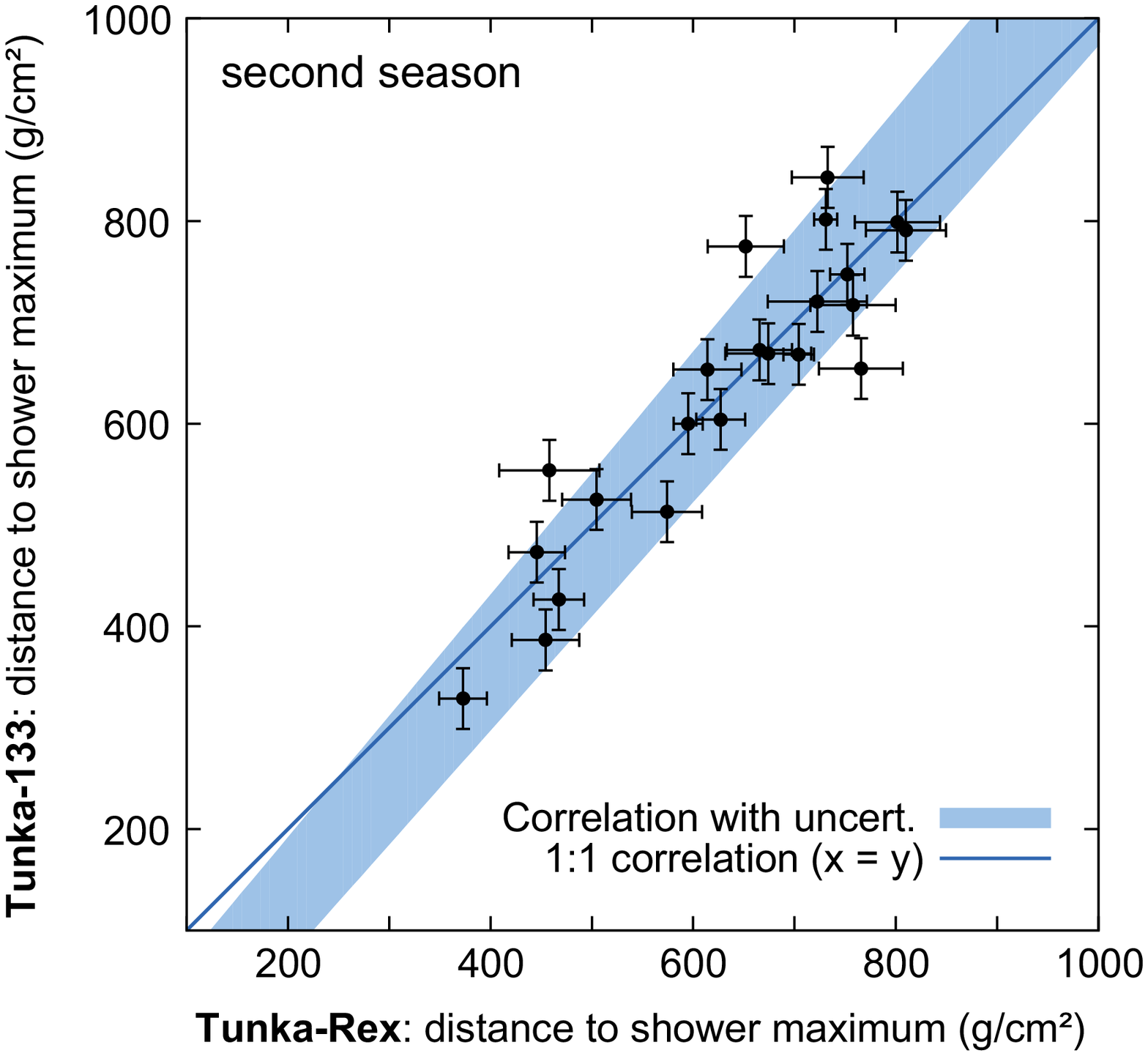}
  \includegraphics[width=0.46\linewidth]{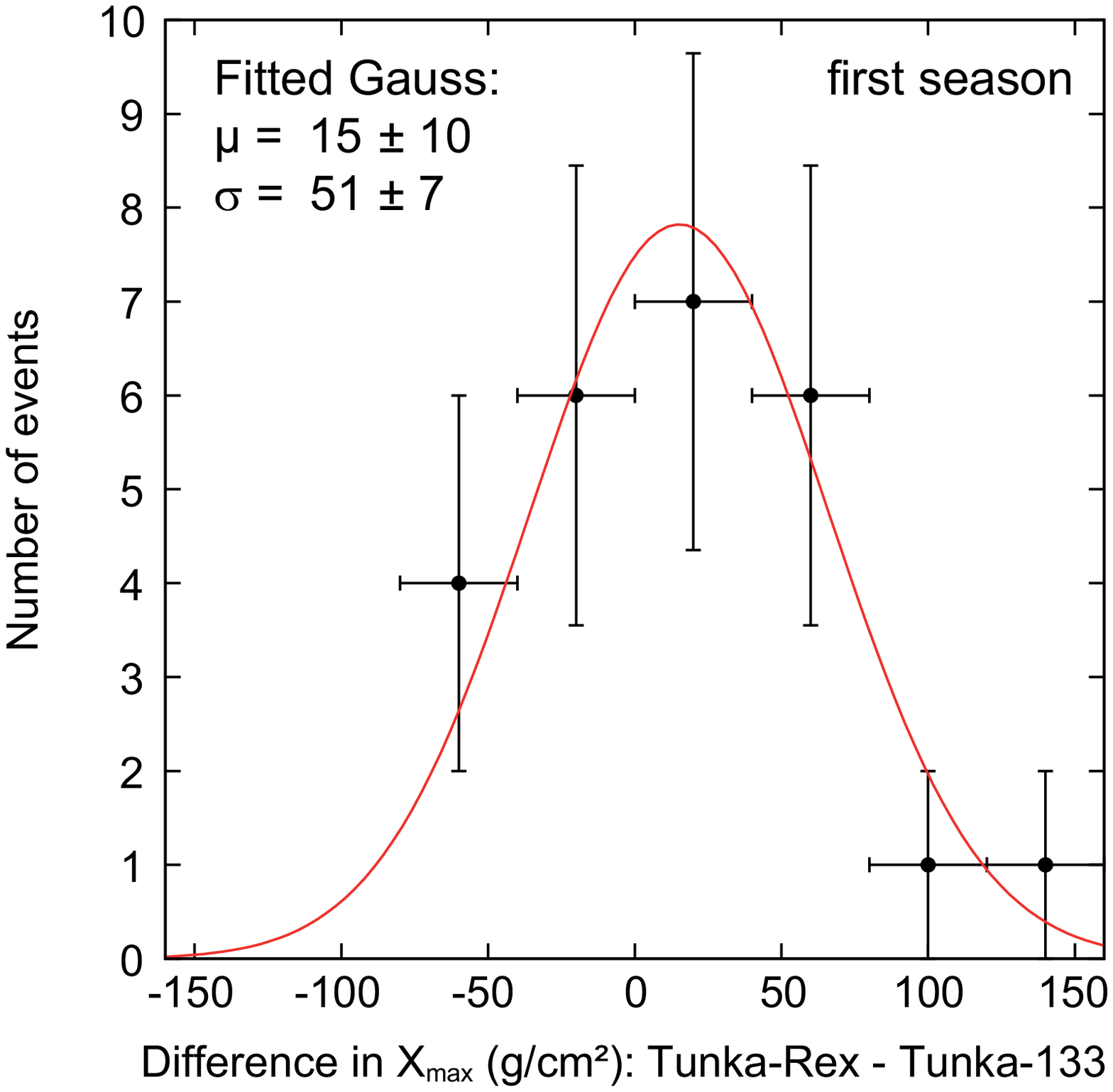}
  \hfill
  \includegraphics[width=0.46\linewidth]{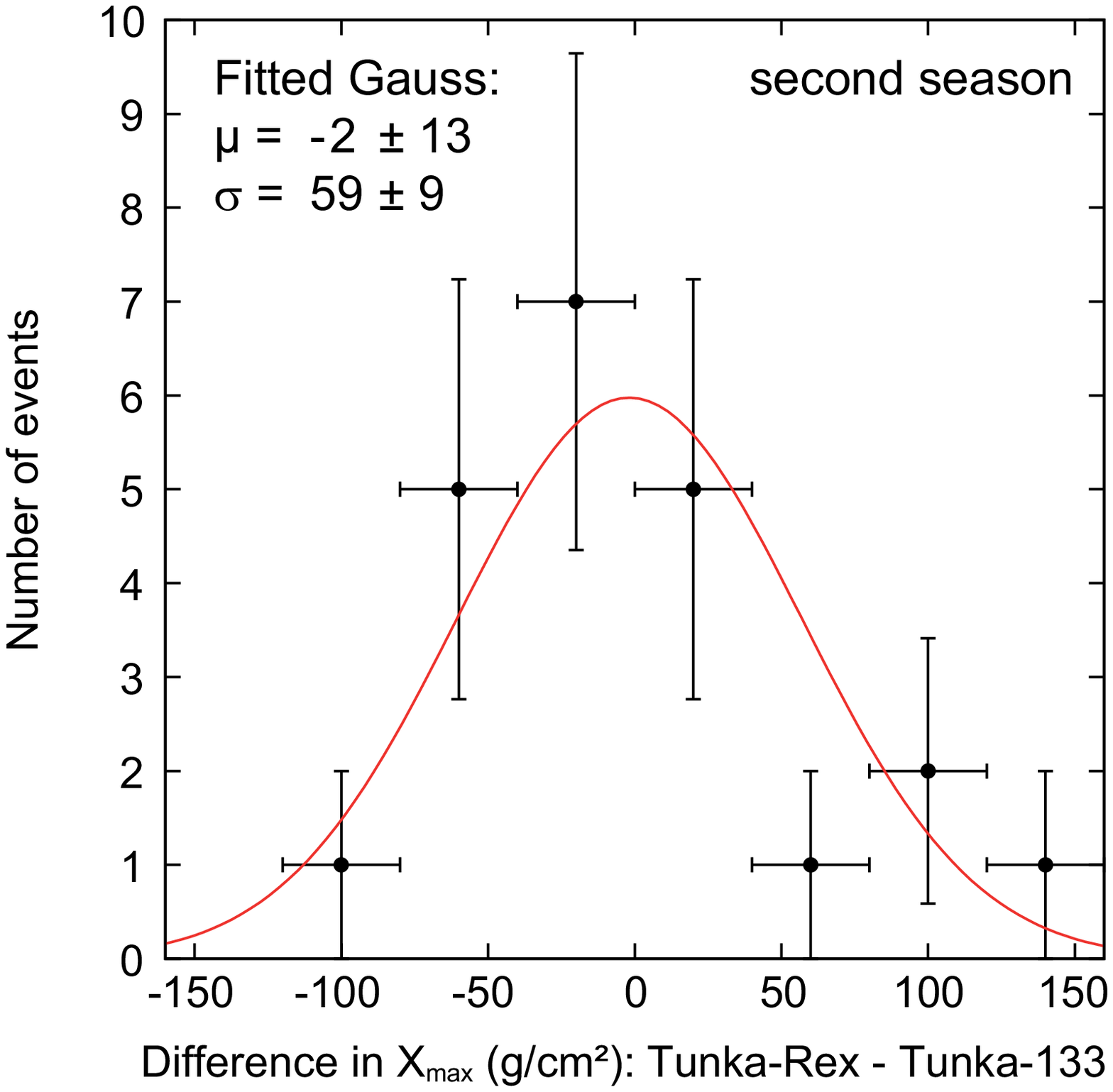}
  \caption{Comparison of the first and second season: distance to shower maximum (cf.~figure~\ref{fig_XmaxCorrelation}).}
  \label{fig_XmaxCorrelationCompareSeasons}
\end{figure*}

\section{Comparison of both seasons}
\label{app_events}
The data set of Tunka-Rex is split in two seasons of about equal size. 
As additional cross-check of the validity of our methods, the Tunka-133 reconstruction of energy and $X_\mathrm{max}$ had been blinded for the second season. 
Only for the first season the Tunka-133 energy and $X_\mathrm{max}$ values had been known, and were used for occasional cross-checks during the development of the Tunka-Rex methods.
After we frozen the Tunka-Rex reconstruction methods, we predicted the energy and $X_\mathrm{max}$ for the second season based on the radio measurements. 
Only afterwards, Tunka-133 unblinded their reconstruction of the second season to us.

The plots in this appendix show the comparison of the Tunka-Rex and Tunka-133 energy and $X_\mathrm{max}$ reconstructions (figures \ref{fig_energyCorrelationCompareSeasons} and \ref{fig_XmaxCorrelationCompareSeasons}) separately for both seasons. 
The results of both seasons are consistent within statistical uncertainties. 
The quality cut on events inside of the inner area ($r < 500\,$m, dashed circle in figure~\ref{fig_map}) was discovered a posteriori by us, i.e., after we had frozen the reconstruction methods for the unblinding procedure. 
Thus, in figure~\ref{fig_XmaxCorrelationCompareSeasons} 5 additional events are present (2 in the first season, 3 in the second season), which are outside of the inner area and slightly degrade the $X_\mathrm{max}$ resolution.

As supplementary material we will upload the list of the events used for the present analysis. 
This list contains the reconstructed energy and $X_{\mathrm{max}}$ values of the events. 
However, the values should not be used for reconstruction of the cosmic-ray energy spectrum or the mass composition.
While unimportant for the present study comparing the Tunka-Rex to the Tunka-133 reconstruction, for such analyses selection biases have to be taken into account, which cannot be derived from the event list alone.

The original event lists for the first \lq tuning\rq~and second \lq prediction\rq~seasons, as created after freezing the reconstruction methods, but before unblinding, are available at:\\
\url{http://www.ikp.kit.edu/tunka-rex/} \\
They can be decrypted using the following Linux command:\\
\texttt{openssl aes-256-cbc -d -in encryptedFile -out decryptedFile}\\
The passwords are \lq TunkaRex1\_H9AoxFywAt\rq~for the tuning season, \\
and \lq TunkaRex2\_z3DumFNfsq\rq~ for the prediction season.\\
All these events are also contained in the above-mentioned list submitted as supplementary material.

\bibliographystyle{JHEP}
\bibliography{crossCalibrationPaper}

\end{document}